\documentclass[11pt]{article}

\usepackage[T1]{fontenc}

\usepackage{amsmath,amssymb,amsthm}
\usepackage[active]{srcltx}

\newcommand{\R}{\mathbb{R}}

\newcommand{\LL}{{\cal L}}

\newcommand{\scal}[2]{\langle #1| #2\rangle}

\newcommand{\eps}{\epsilon}
\newcommand{\veps}{\varepsilon}
\newcommand{\ot}{\otimes}

\newcommand{\bld}[1]{\boldsymbol{#1}}
\newcommand{\tl}[1]{\tilde{#1}}
\newcommand{\we}{\wedge}
\newcommand{\un}[1]{\underline{#1}}
\newcommand{\bth}{\boldsymbol{\theta}}
\newcommand{\vth}{\vec{\theta}}
\newcommand{\lr}{\lrcorner}
\newcommand{\ubth}{\underline{\boldsymbol{\theta}}}
\newcommand{\bed}{\boldsymbol{d}}
\newcommand{\Star}[1]{*'\!\!_{#1}\,}

\newcounter{mnotecount}[section]

\numberwithin{equation}{section}
\numberwithin{thr}{section}

\begin{document}

\title{Hamiltonian formulation of a simple theory of the teleparallel geometry\footnote{This is an author-created version of an article accepted for publication in Classical and Quantum Gravity, including the subsequent correction of a linguistic error. IOP Publishing Ltd is not responsible for any errors or omissions in this version of the manuscript or any version derived from it. The definitive publisher authenticated version is available online at http://dx.doi.org/10.1088/0264-9381/29/4/045008.}}
\author{ Andrzej Oko{\l}\'ow, J\k{e}drzej \'Swie\.zewski}
\date{February 15, 2022}

\maketitle
\begin{center}
{\it  Institute of Theoretical Physics, Warsaw University\\ ul. Ho\.{z}a 69, 00-681 Warsaw, Poland\smallskip\\
oko@fuw.edu.pl\\
swiezew@fuw.edu.pl}
\end{center}
\medskip

\begin{abstract}
A theory of cotetrad fields on a four-dimensional manifold is considered. Its configuration space coincides with that of the Teleparallel Equivalent of General Relativity but its dynamics is much simpler. We carry out the Legendre transformation and derive a Hamiltonian and a constraint algebra. 
\end{abstract}

\section{Introduction}

There are many distinct formulations of General Relativity (GR) differing from each other by the way one encodes information about the gravitational field. The original approach by Einstein uses a spacetime metric as the fundamental variable \cite{ein-1}. In the so-called Palatini formulation the basic variables are a cotetrad field on a spacetime and a connection one-form of non-zero curvature \cite{ein-2}. GR can be viewed as a constrained $BF$-theory where the basic variables are a connection one-form and a two-form \cite{pleb,capo}. Another formulation is the Teleparallel Equivalent of GR (TEGR) (see e.g. \cite{arcos,obh,mielke} and references therein) where the fundamental variables are either a cotetrad field and a connection of zero curvature or just a cotetrad field.

Endeavoring to formulate a quantum model of gravity one can choose a quantization method and try to apply it to any of different formulations of GR. There is of course no guarantee that a particular method will work well with a chosen formulation, but examples of Loop Quantum Gravity (LQG) (see e.g. \cite{BIQ,Thmnn}) and Spin Foams (SF) (see e.g. \cite{ale}) show that it is possible---LQG is a result of canonical quantization applied to the formulation of GR in terms of the Ashtekar-Barbero connection, while SF come from the $BF$-like approach to GR subjected to (appropriately defined) path integral method.  

We are concerned with applying a canonical quantization to TEGR formulated as a theory of coframes (cotetrad fields). Since Hamiltonian formulations of this version of TEGR\footnote{A Hamiltonian description of TEGR formulated as a theory of coframes and connections can be found in \cite{bl}.} appear to be rather complicated \cite{nester,maluf,maluf-1,maluf-2} it seems reasonable to check first whether canonical quantization can be successfully applied to a theory which shares the configuration space with TEGR but differs from it by having much simpler dynamics. In this paper we present a Hamiltonian formulation of such a theory as a first step towards its quantum model.

The theory will be formulated in terms of differential forms---an action defining the dynamics of the theory will be given as an integral of a four-form built from a coframe by means of the exterior derivative, the wedge product and a Hodge operator. The Hamiltonian framework of the theory will be also expressed in terms of forms. It turns out that it is possible to describe a Hamiltonian framework of TEGR in the same fashion \cite{oko} and the research on the theory  presented in this paper was a preparatory exercise for that. 

The paper is organized as follows: after short preliminaries (Section 2) we define the theory (Section 3). Next, in Section 4 we describe a $3+1$ decomposition of all objects needed to define the action and, finally, of the action itself. In Section 5 the Legendre transformation is carried out and a Hamiltonian is derived. In Section 6 we derive a constraint algebra and in Section 7 we present a short discussion of the results obtained. In Appendix placed after Section 7 we recall shortly some basic definitions, prove many formulae applied in the paper and present a general Hamiltonian formalism adapted to differential forms based on \cite{ham-diff,mielke}.

\section{Preliminaries}

Let $\mathbb{M}$ be a four-dimensional oriented vector space equipped with a scalar product $\eta$ of signature\footnote{According to the definition of signature given in Appendix \ref{vol} the signature of $\eta$ is $1$.} $(-,+,+,+)$. We fix an orthonormal basis $(v_A)$ $(A=0,1,2,3)$ such that the components $(\eta_{AB})$ of $\eta$ given by the basis form the matrix ${\rm diag}(-1,1,1,1)$. The matrix $(\eta_{AB})$ and its inverse $(\eta^{AB})$ will be used to, respectively, lower and raise capital Latin letter indeces.     

Let $\cal M$ be a four-dimensional oriented manifold. We assume that there exists a smooth  map $\bth:T{\cal M}\to \mathbb{M}$ such that for every $y\in{\cal M}$ the restriction of $\bth$ to the tangent space $T_y{\cal M}$ is a linear isomorphism between the tangent space and $\mathbb{M}$ which preserves the orientations. The map $\bth$ can be expressed by means of the orthogonal basis $(v_A)$ as
\[
\bth=\bth^A\ot v_A,
\]           
where $(\bth^A)$ are one-forms on $\cal M$. Obviously, the one-forms $(\bth^A)$ form a {\em coframe} or a {\em cotetrad field} on the manifold. If $(x^\mu)$, ($\mu=0,1,2,3$), is a local coordinate frame on $\cal M$ compatible with its orientation then the determinant of the matrix $(\bth^A_\mu)$ built form the components of the forms $\bth^A$ in the coordinate frame is positive,
\begin{equation}
\det(\bth^A_\mu)>0.
\label{t-det}
\end{equation}

The map $\bth$ can be used to pull back the scalar product $\eta$ on $\mathbb{M}$ to a metric on the manifold $\cal M$ turning thereby the manifold into a {\em spacetime}. We will denote the resulting metric by $g$,
\begin{equation}
g:=\eta_{AB}\bth^A\ot\bth^B.
\label{g}
\end{equation}
The metric $g$ defines\footnote{We recall the definitions in Appendices \ref{vol} and \ref{Hodge-app}.} a volume form $\bld{\eps}$ on $\cal M$ and a Hodge operator $\star$ mapping differential $k$-forms to $(4-k)$-forms on the manifold $(k=0,1,2,3,4)$.      

\section{Definition of the theory}

In this paper we will consider a theory of cotetrad fields on $\cal M$ which means that the configuration space of the theory will be a set of all the maps $\bth$ which satisfy the assumptions listed in the previous section. Since we are interested in a simple toy-model the dynamics of the theory will be given by the following action \cite{itin}:  
\begin{equation}
S[\bth^A]=-\frac{1}{2}\int\bed\bth^A\we\star\bed\bth_A,
\label{act}
\end{equation}
which seems to be the simplest background independent action quadratic in derivatives of $\bth^A$. Recall that the Hodge operator $\star$ is given by the metric $g=\eta_{AB}\bth^A\ot\bth^B$ and therefore is a function of $\bth^A$. The action is invariant with respect to
\begin{enumerate}
\item diffeomorphisms of $\cal M$: $S[\varphi^*\bth^A]=S[\bth^A]$, where $\varphi^*$ denotes the pull-back given by a diffeomorphism $\varphi$ on $\cal M$,
\item {\em global} Lorentz transformations: $S[\Lambda^A{}_B\bth^B]=S[\bth^A]$, where $\Lambda^A{}_B$ is a {\em constant} matrix belonging to the Lorentz group.
\end{enumerate}

Let us now describe a relation between the theory just defined and TEGR. 
As mentioned in the introduction TEGR can be formulated as a theory of cotetrad fields, which means that both theories under considerations share the same configuration space. Now let us compare actions defining the theories.

Consider a linear space of  two-forms on $\mathbb{M}$ valued in $\mathbb{M}$. Such a two-form is of the following form 
\[
\Phi=\Phi^A\ot v_A=\frac{1}{2}\Phi^A{}_{BC}\tl{v}^B\we\tl{v}^C\ot v_A,
\]   
where $\Phi^A{}_{BC}=-\Phi^A{}_{CB}$ and $(\tl{v}^A)$ is a basis dual to $(v_A)$. There is a natural representation of the Lorentz group on this linear space:
\[
\Phi^A{}_{BC}\mapsto\Lambda^{A}{}_{A'}\Lambda^{-1B'}{}_B\Lambda^{-1C'}{}_C\Phi^{A'}{}_{B'C'},
\]
where $\Lambda^A{}_B$ is a Lorentz matrix. This representation acts pointwise on the two-form $\bed\bth^A=\frac{1}{2}(\bed \bth^A)_{BC}\bth^B\we\bth^C$ and provides a decomposition of $\bed\bth^A$ into irreducible components \cite{mccrea,hehl}
\begin{equation}
\begin{aligned}
^{(1)}\!\bed\bth^A&:=\bed\bth^A-^{(2)}\!\bed\bth^A-^{(3)}\!\bed\bth^A,\\
^{(2)}\!\bed\bth^A&:=\frac{1}{3}\bth^A\we(\bld{e}_B\lr \bed\bth^B),\\
^{(3)}\!\bed\bth^A&:=\frac{1}{3}\bld{e}^A\lr(\bth_B\we \bed\bth^B),
\end{aligned} 
\label{Phi-dec}
\end{equation}
where $(\bld{e}^B)$ is a reper dual to $(\bth^A)$, and $\lr$ denotes a contraction of a vector field with a differential form\footnote{Let $\alpha$ be a differential $k$-form and $X$ a vector field on a manifold. Then $X\lr\alpha$ is a $(k-1)$-form such that for any vector fields $X_1,\ldots,X_{k-1}$ 
\[
(X\lr\alpha)(X_1,\ldots,X_{k-1}):=\alpha(X,X_1,\ldots,X_{k-1}).
\]      
}. Using this decomposition one can define a family of actions quadratic in $\bed\bth^A$ \cite{mielke}:
\begin{equation}
S[\bth^A;a_1,a_2,a_3]:=-\frac{1}{2}\int \bed \bth^A\we\star(\sum_{i=1}^3 a_i\,^{(i)}\! \bed\bth_A),
\label{aaa-act}
\end{equation}
where $\{a_i\}$ are real numbers. Setting $a_1=1$, $a_2=-2$ and $a_3=-1/2$ one obtains an action of TEGR, while setting $a_1=a_2=a_3=1$ one arrives at \eqref{act}. 
Note that the action \eqref{act} is (modulo a constant factor) the simplest action among \eqref{aaa-act}---in this case all the $\{a_i\}$ are equal and consequently the irreducible components $\{^{(i)}\! \bed\bth^A\}$, being quite complicated functions of $\bth^A$, sum up to $\bed\bth^A$.      

Alternatively, an action of TEGR can be expressed as follows \cite{waldyr}:
\[
S[\bth^A]=-\frac{1}{2}\int\bed\bth^A\we\star\bed\bth_A-(\star d\star \bth^A)\we d\star\bth_A-\frac{1}{2}(\bed\bth^A\we\bth_A)\we\star(\bed\bth^B\we\bth_B).
\]
Omitting the last two terms one gets \eqref{act}. 

On the other hand the action above can be rewritten as \cite{waldyr}:
\begin{equation}
S[\bth^A]=\int-\frac{1}{2}(\bed\bth^A\we\bth_B)\we\star(\bed\bth^B\we\bth_A)+\frac{1}{4}(\bed\bth^A\we\bth_A)\we\star(\bed\bth^B\we\bth_B).
\label{tegr-act}
\end{equation}
Note that there is a similarity between this action and \eqref{act}---the integrand of the latter one is just ``a square'' of $\bed\bth^A$ (defined by the Hodge operator $\star$ and the scalar product $\eta$), while the integrand of the former one consists of two ``squares'' of $\bed\bth^A\we\bth^B$. Due to this similarity the Hamiltonian analysis presented in this paper turned out to be very helpful while studying the Hamiltonian structure of TEGR based on the action \eqref{tegr-act} \cite{oko}.      

\section{$3+1$ decomposition \label{3+1}}

To carry out a $3+1$ decomposition of the action \eqref{act} we have to impose some additional assumptions on the manifold $\cal M$ and the map $\bth$. We require that 
\begin{enumerate}
\item ${\cal M}=\R\times\Sigma$, where $\Sigma$ is a three dimensional manifold.
\end{enumerate}
This assumption allows us to introduce a family of curves in $\cal M$ parameterized by points of $\Sigma$---given $x\in\Sigma$ we define
\begin{equation}
\R\ni t\mapsto (t,x)\in\R\times\Sigma={\cal M}.
\label{curv}
\end{equation}
These curves generates a global vector field on $\cal M$ which will be denoted by $\partial_t$. We require moreover that
\begin{enumerate}
\setcounter{enumi}{1}
\item the map $\bth$ is such that $\partial_t$ is timelike with respect to the metric $g$ defined by $\bth$.     
\item the map $\bth$ is such that for every $t\in\R$ the submanifold $\Sigma_t:=\{t\}\times\Sigma$ is spatial with respect to $g$. 
\end{enumerate}
Now we can use the vector field $\partial_t$ to define a time orientation of $\cal M$---by definition $\partial_t$ is future directed.   

In order to not be troubled by boundary terms in the Hamiltonian formulation we assume that
\begin{enumerate}
\setcounter{enumi}{3}
\item $\Sigma$ is a compact manifold without boundary. 
\end{enumerate}

Assumption 1 allows us to define a function on $\cal M$ which maps a point $y$ to a number $t$  such that $y\in\Sigma_t$. Abusing the notation we will use the letter $t$ to denote the function.  Let $(x^i)$, $(i=1,2,3)$, be local coordinates on $\Sigma$. The coordinates together with the function $t$ define local coordinates on $\cal M$ which associate with an appropriate $y\in\Sigma_t$ four numbers $(x^0\equiv t,x^i)\equiv(x^\mu)$. Obviously, on the domain of such a coordinate frame the vector field $\partial_0$ given by the frame coincides with $\partial_t$ generated by the curves \eqref{curv}. Since now we will restrict ourselves to coordinate frames $(x^\mu)$ on $\cal M$ of this sort assuming additionally that each frame we are going to use is compatible with the orientation of the manifold. 

Note that these coordinate frames induce an orientation of $\Sigma$ which since now will be treated as an {\em oriented} manifold. 

Let us finally emphasize that in this paper the spacetime indeces will be denoted by lower case Greek letters and will range from $0$ to $3$ and the spatial indeces will be denoted by lower case Latin letters and will range from $1$ to $3$.    

Now we are ready to carry out a $3+1$ decomposition of all relevant objects: differential forms,  the coframe $(\bth^A)$, the metric $g$, the volume form $\bld{\eps}$, the Hodge operator $\star$ and finally the action \eqref{act}.        

\subsection{Decomposition of differential forms \label{dec-forms}}

Denote by $\bed$ the exterior derivative of forms on $\cal M$ and by $d$ the exterior derivative of forms on $\Sigma$. A $k$-form $\alpha$ on $\cal M$ can be decomposed with respect to the decomposition ${\cal M}=\R\times\Sigma$ as follows \cite{ham-diff}
\[
\alpha={}^\perp\alpha+\un{\alpha},
\]    
where
\[
{}^\perp\alpha:=\bed t\we \alpha_\perp, \ \ \alpha_\perp:=\partial_t\lr\alpha,
\]
is its ``timelike'' part and
\[
\un{\alpha}:=\partial_t\lr(\bed t\we\alpha)
\]
its ``spatial'' part. 

$\un{\alpha}$ is a form on $\cal M$ which can be expressed in a coordinate frame $(t,x^i)$ as
\[
\un{\alpha}=\frac{1}{k!}\alpha_{i_1\ldots i_k}(t,x^i)\bed x^{i_1}\we\ldots\we\bed x^{i_k}.
\]  
The form naturally defines a form on $\Sigma$ (or more precisely, a one parameter family of forms on $\Sigma$ the parameter being the coordinate $t$) 
\[
\un{\alpha}':=\frac{1}{k!}\alpha_{i_1\ldots i_k}(t,x^i)d x^{i_1}\we\ldots\we dx^{i_k}.
\]
Moreover, it is possible to restore the original form $\un{\alpha}$ from $\un{\alpha}'$: given the latter one we define
\[
\partial_t\lr\un{\alpha}:=0, \ \ \ \un{\alpha}(\vec{X}_1,\ldots,\vec{X}_k):=\un{\alpha}'(\vec{X}_1,\ldots,\vec{X}_k), 
\]    
for all vector fields $(\vec{X}_1,\ldots,\vec{X}_k)$ tangent to the foliation $\{\Sigma_t\}$ of $\cal M$. Therefore in the sequel we will not distinguish between $\un{\alpha}$ and $\un{\alpha}'$. There is, however, one subtlety concerning Lie derivatives of $\un{\alpha}$ and $\un{\alpha}'$. Let $\vec{X}$ be a vector field on $\cal M$ tangent to the foliation $\{\Sigma_{t}\}$. Denote by $\LL_{\vec{X}}$ the Lie derivative on $\cal M$ with respect to $\vec{X}$  and by $\LL'_{\vec{X}}$  the Lie derivative on $\Sigma_t$ with respect to a restriction of $\vec{X}$ to the submanifold. Then in general $\LL_{\vec{X}}\un{\alpha}$ cannot be identified with $\LL'_{\vec{X}}\un{\alpha}'$. Indeed, if $\alpha$ is a one-form on $\cal M$ then    
\[
\LL_{\vec{N}}\un{\alpha}=(N^\mu\partial_\mu\un{\alpha}_\nu+\un{\alpha}_\mu\partial_\nu N^\mu)\bed x^\nu=\un{\alpha}_i\partial_0 N^i \bed t+(N^i\partial_i\un{\alpha}_j+\un{\alpha}_i\partial_j N^i)\bed x^j.
\]
and only the last term in this equation can be identified with $\LL'_{\vec{X}}\un{\alpha}'$. However, in the sequel we will never encounter Lie derivatives $\LL_{\vec{X}}\un{\alpha}$ as defined above, but we will do encounter derivatives $\LL'_{\vec{X}}\un{\alpha}'$. Since we would like our notation to be as simple as possible, since now we will use the symbol $\LL_{\vec{X}}\un{\alpha}$ to denote the derivative $\LL'_{\vec{X}}\un{\alpha}'$.

Similarly, $\alpha_\perp$ is a form on $\cal M$, but it can be treated as a (one parameter family of) form(s) on $\Sigma$.   

Here we present a list of basic properties of the maps $\alpha\mapsto\alpha^\perp$ and $\alpha\mapsto\un{\alpha}$ \cite{ham-diff,mielke}:
\begin{equation}
\begin{aligned}
&{}^\perp({}^\perp\alpha)={}^\perp\alpha, & &\un{({}^\perp\alpha)}={}^\perp(\un{\alpha})=0,  & &\un{(\un{\alpha})}=\un{\alpha},\\
&{}^\perp(\alpha\we\beta)=({}^\perp\alpha)\we\un{\beta}+\un{\alpha}\we({}^\perp\beta), & & \un{(\alpha\we\beta)}=\un{\alpha}\we\un{\beta}, & & \un{\alpha_\perp}=\alpha_\perp,\\
& \partial_t\lr\un{\alpha}=0, & & (\bed\alpha)_\perp=\LL_{\partial_t}\un{\alpha}-d\alpha_\perp, & & \un{(\bed \alpha)}=d\un{\alpha},\\
& \bed \alpha=\bed t\we\LL_{\partial_t}\un{\alpha}-\bed t\we d\alpha_\perp+d\un{\alpha} & &   & & 
\end{aligned}
\label{perp-un}  
\end{equation}
where $\LL_{\partial_t}$ denotes the Lie derivative with respect to the vector field $\partial_t$. 

\subsection{Decomposition of the coframe}

Since each $\bth^A$ is a one-form it decomposes as
\begin{equation}
\bth^A=\bth^A_\perp \bed t + \ubth^A.
\label{corep-dec}
\end{equation}
Our goal now is to express $\bth^A_\perp$ as a function of $\ubth^A$ and some additional parameters.

To this end we are going to show that there exists a function $\xi^A$ on $\cal M$ valued in $\mathbb{M}$  such that \cite{nester}
\begin{equation}
\xi^A\ubth_A=0 \ \ \ \text{and} \ \ \ \xi^A\xi_A=-1.
\label{xi-df}
\end{equation}
The first condition allows us to conclude that    
\[
\xi^A\propto\veps^A{}_{BCD}\bth^B_1\bth^C_2\bth^D_3
\]
where $\veps_{ABCD}$ is a volume form on $\mathbb{M}$ defined by the scalar product $\eta$. However, this expression turns out not to be very convenient, so let us change it a little bit. Denote by $q$ an Euclidean metric induced on $\Sigma_t$ by the metric $g$: 
\begin{equation}
\begin{gathered}
q=q_{ij}dx^i\ot dx^j:=g_{ij}dx^i\ot dx^j=\eta_{AB}\ubth^A\ot\ubth^B,\\
q_{ij}=\eta_{AB}\bth^A_i\bth^B_j.
\end{gathered}
\label{q}
\end{equation}   
The metric $q$ defines a volume form $\eps$ on $\Sigma_t$ and a Hodge operator $*$ acting on the forms on the submanifold. Let $\eps^{ijk}$ be the components of a tensor obtained from $\eps$ by raising its indeces with a metric inverse to $q$. Then
\[
\xi^A=\kappa\veps^A{}_{BCD}{\eps}^{ijk}\bth^B_i\bth^C_j\bth^D_k=\kappa\veps^A{}_{BCD}*(\ubth^B\we\ubth^C\we\ubth^D),
\]       
where $\kappa$ is a function. Now using the identities 
\begin{align}
\veps^{AB'C'D'}\veps_{ABCD}&=-3!\delta^{[B'}{}_B\delta^{C'}{}_C\delta^{D']}{}_D,\label{ee-4}\\
\eps^{i'j'k'}\eps_{ijk}&=3!\delta^{[i'}{}_i\delta^{j'}{}_j\delta^{k']}{}_k \nonumber
\end{align}
being particular cases of a general identity \eqref{eps-eps!} we impose on $\xi^A$ the second condition in \eqref{xi-df} obtaining thereby 
\begin{equation}
\xi^A=\pm\frac{1}{3!}\veps^A{}_{BCD}*(\ubth^B\we\ubth^C\we\ubth^D).
\label{xi}
\end{equation}

Using a coordinate frame $(t\equiv x^0,x^i)$ we see that
\[
\bth^A_\perp=\bth^A_0 \ \ \ \text{and} \ \ \ \ubth^A=\bth^A_idx^i.
\] 
Fixing the value of the index $\mu$ we can treat $\bth^A_\mu$ as a function valued in $\mathbb{M}$. The conditions \eqref{xi-df} mean that for every $y\in{\cal M}$ the vectors $(\xi^A(y),\bth^A_i(y))$ form a basis of $\mathbb{M}$. Therefore there exists a function $N$ and a vector field $\vec{N}=N^i\partial_i$ on ${\cal M}$ tangent to the foliation $\{\Sigma_t\}$ such that \cite{nester}
\begin{equation}
\bth^A_\perp=N\xi^A+N^i\bth^A_i=N\xi^A+\vec{N}\lr\ubth^A.
\label{t-perp-dec}
\end{equation}
The function $N$ will be called {\em lapse} and the vector field $\vec{N}$ will be called {\em shift}.
 
Let us now comment on the result \eqref{xi}. It is ambiguous because the second condition in \eqref{xi-df} is quadratic in $\xi^A$. This is, however, not a serious problem since  in \eqref{t-perp-dec} $\xi^A$ is multiplied by $N$ and any change of the sign of $\xi^A$ can be absorbed by a change of the sign of $N$. Nevertheless, it would be convenient to fix the sign in \eqref{xi}. Before we will do this let us first apply the condition \eqref{t-det} (recall that the coordinate frame $(t,x^i)$ is assumed to be compatible with the orientation of $\cal M$):
\begin{multline*}
\det(\bth^A_\mu)=\bth^A_0\bth^B_1\bth^C_2\bth^D_3\veps_{ABCD}=\\=N\xi^A\bth^B_1\bth^C_2\bth^D_3\veps_{ABCD}=\pm\frac{N}{3!}\veps^A{}_{B'C'D'}\veps_{ABCD}{\eps}^{ijk}\bth^{B'}_i\bth^{C'}_j\bth^{D'}_k\bth^B_1\bth^C_2\bth^D_3=\\=\mp N \eps^{ijk}q_{i1}q_{j2}q_{k3}=\mp N\eps_{123}=\mp N\sqrt{\det q_{ij}}>0, 
\end{multline*}
where we used \eqref{ee-4} and \eqref{q}. This result means that either $(i)$ the sign in \eqref{xi} is $-$ and $N>0$ or $(ii)$ the sign is $+$ and $N<0$. To remove the ambiguity we choose $(i)$. 

Let us summarize this subsection by expressing the final form of the decomposition of the function $\bth^A_\perp$: 
\begin{align}
&\bth^A_\perp=N\xi^A+\vec{N}\lr\un{\bth}^A, 
\label{theta-0-xi-a}\\
&\xi^A=-\frac{1}{3!}\veps^A{}_{BCD}*(\un{\bth}^B\we\un{\bth}^C\we\un{\bth}^D)
\label{xi-a},\\
&N>0. \label{N>0}
\end{align} 

\subsection{Decomposition of metric and volume form}

Here we will use the results of the previous subsection to obtain a 3+1 decomposition of the metric $g$ defined by \eqref{g}. Note first that we already introduced (see \eqref{q}) the spatial metric $q$ on $\Sigma$ as the one induced on the manifold by the space-time metric $g$. The metric $q$ and its inverse $q^{-1}$,
\begin{equation}
q^{-1}:=q^{ij}\partial_i\ot\partial_j, \ \ \ q^{ij}q_{jk}=\delta^i{}_k,
\label{q-1}
\end{equation}
will be used to, respectively, lower and raise indeces of tensors defined on $\Sigma$. 
As mentioned earlier the metric $q$ defines the volume form $\eps$ on $\Sigma$ and the Hodge operator $*$. Let us emphasize that (as it follows from \eqref{q}) the metric $q$ can be defined explicitely in terms of the restricted forms $(\un{\bth}^A)$. Therefore all object defined by $q$ (as $q^{-1}$, $\eps$ and $*$) are in fact functions of $(\un{\bth}^A)$.      

It is clear that
\begin{equation}
g_{\mu\nu}=\eta_{AB}\bth^A_\mu\bth^B_\nu,
\label{g-comp}
\end{equation}
hence by applying \eqref{theta-0-xi-a} and \eqref{q} we get
\begin{equation}
g_{00}=-N^2+N^iN^jq_{ij}=-N^2+N^iN_i, \ \ \ g_{0i}=N^jq_{ji}=N_i, \ \ \ g_{ij}=q_{ij},
\label{g-dec-comp}
\end{equation}
which is the standard 3+1 decomposition of the metric \cite{adm}. Equivalently, 
\begin{equation}
g=(-N^2+N^iN_i)\,\bed t^2+2N_i\,\bed t\bed x^i+q.
\label{g-dec}
\end{equation}

Denote by $T$ a {\em future directed} timelike vector field orthogonal at every point $x\in{\cal M}$ to $\Sigma_t$ passing through $x$ and normed, i.e., $g(T,T)=-1$. Then $(T,\partial_i)$ is a (local) reper on $\cal M$, hence there exist four numbers $n>0$ and $(n^i)$ such that     
\[
\partial_t=nT+n^i\partial_i.
\]         
Using this formula to calculate $g_{00}$ and $g_{0i}$ and comparing the results with \eqref{g-dec-comp} we see that $n=N$ and $n^i=N^i$, that is,
\[
\partial_t=NT+\vec{N}
\]      
which justify calling $N$ the lapse and $\vec{N}$ the shift. 

Later on we will need a 3+1 decomposition of the metric $g^{-1}$ inverse to $g$ and the volume form $\bld{\eps}$ on $\cal M$. Let  
\begin{equation}
\tilde{T}=N\bed t, \ \ \ \tilde{T}^i=N^i\bed t+\bed x^i.
\label{TT}
\end{equation}
Then
\begin{equation}
g=-\tilde{T}^2+q_{ij}\tilde{T}^i\ot\tilde{T}^j
\label{g-TT}
\end{equation}
The vector fields $(T,\partial_i)$ turn out to form a tetrad dual to the cotetrad $(\tilde{T},\tilde{T}^i)$ hence \cite{adm}
\begin{equation}
g^{-1}=-T^2+q^{ij}\partial_i\ot\partial_j=\frac{1}{N^2}\Big(-\partial_t\ot\partial_t+2N^i\partial_t\ot\partial_i+(N^2q^{ij}-N^iN^j)\partial_i\ot\partial_j\Big).
\label{g-1-dec}
\end{equation}

It follows from \eqref{g-TT} and \eqref{TT} that the volume form $\bld{\eps}$ decomposes as follows 
\begin{equation}
\bld{\eps}=\sqrt{\det q_{ij}}\,\tilde{T}\we\tilde{T}^1\we\tilde{T}^2\we\tilde{T}^3=N\bed t\we(\sqrt{\det q_{ij}}\,\bed x^1\we \bed x^2\we \bed x^3)=N \bed t\we {\eps},
\label{ee-dec}
\end{equation}
where ${\eps}$ is the volume form on $\Sigma_t$ defined by $q$.

\subsection{Decomposition of $\alpha\we\star \beta$ }

Let $\alpha,\beta$ be $k$-forms on $\cal M$. Then 
\begin{equation}
\alpha\we\star\beta=-N^{-1}\bed t\we(\alpha_\perp-\vec{N}\lrcorner\un{\alpha})\we{*}(\beta_\perp-\vec{N}\lrcorner\un{\beta})+N\,\bed t\we\un{\alpha}\we{*}\,\un{\beta},
\label{A*B-dec}
\end{equation}
where ${*}$ is the Hodge operator defined by the metric $q$ on $\Sigma_t$.

The decomposition \eqref{A*B-dec} is proven in Appendix \ref{Hodge-app} in a general case, i.e., for any $k$-forms $\alpha,\beta$ on an $n$-dimensional manifold equipped with a Lorentzian metric $g$. 

Let us note as a digression that \eqref{A*B-dec} allows us to express the parts ${}^\perp(\star\beta)$ and $\un{\star\beta}$ by means of $\beta_\perp$, $\un{\beta}$, the lapse, the shift, the operator $*$ and the one-form $\bed t$. To show this we assume for a while that $\dim{\cal M}=n$. Then $\alpha\we\star\beta$ is an $n$-form and therefore $\un{\alpha\we\star\beta}=0$. Consequently, by virtue of \eqref{perp-un}
\[
\alpha\we\star\beta={}^\perp(\alpha\we\star\beta)={}^\perp\alpha\we\un{\star\beta}+\un{\alpha}\we{}^\perp(\star\beta).
\]  
On the other hand, \eqref{A*B-dec} can be transformed to the following form:
\[
\alpha\we\star\beta={}^\perp\alpha\we\Big(-N^{-1}*(\beta_\perp-\vec{N}\lr\un{\beta})\Big)+\un{\alpha}\we\Big(-\bed t\we N^{-1}\vec{N}\lr*(\beta_\perp-\vec{N}\lr\un{\beta})+(-1)^kN\bed t\we*\un{\beta}\Big).
\]
Note now that in the equation above ${}^\perp\alpha$ is multiplied by a ``spatial'' form while $\un{\alpha}$ by a ``timelike'' form. Moreover, the last two equations hold for every $\alpha$. Taking into account the fact that ${}^\perp\alpha$ and $\un{\alpha}$ are independent we obtain   
\begin{align*}
\un{\star\beta}&=-N^{-1}*(\beta_\perp-\vec{N}\lr\un{\beta}),\\
{}^\perp(\star\beta)&=-\bed t\we N^{-1}\vec{N}\lr*(\beta_\perp-\vec{N}\lr\un{\beta})+(-1)^kN\bed t\we*\un{\beta}.
\end{align*}
However, we will not use these two formulae in this paper. 

\subsection{Decomposition of the action \label{act-dec-ss}}

According to \eqref{A*B-dec} the action \eqref{act} can be decomposed as follows
\[
S[\bth^A]=\int \frac{1}{2N}\bed t\we((\bed \bth^A)_\perp-\vec{N}\lrcorner\un{\bed \bth}^A)\we{*}((\bed \bth_{A})_{\perp}-\vec{N}\lrcorner\un{\bed \bth}_A)-\frac{N}{2}\bed t\we\un{\bed \bth}^A\we{*}\un{\bed \bth}_A,
\]   
where at this moment $N$ and $\vec{N}$ are functions of $\bth^A$ (an explicite form of these functions can be obtained from \eqref{g-comp}, \eqref{g-dec-comp} and \eqref{N>0}). Expressing $(\bed\bth^A)_\perp$ by means of $\bth^A_\perp$ and $\un{\bth}^A$ as shown in \eqref{perp-un} we rewrite the action in the following form
\begin{multline}
S[\bth^A_\perp,\un{\bth}^B]=\int \frac{1}{2N}\bed t\we(\LL_{\partial_t}\un{\bth}^A-{d}(\bth^A_\perp)-\vec{N}\lrcorner {d}\un{\bth}^A)\we{*}(\LL_{\partial_t}\un{\bth}_A-{d}(\bth_{A\perp})-\vec{N}\lrcorner{d}\un{\bth}_A)-\\-\frac{N}{2}\bed t\we {d}\un{\bth}^A\we{*}{d}\un{\bth}_A=\\=\int \frac{1}{2N}\bed t\we(\LL_{\partial_t}\un{\bth}^A-E^A)\we{*}(\LL_{\partial_t}\un{\bth}_A-E_A)-\frac{N}{2}\bed t\we{d}\un{\bth}^A\we{*}{d}\un{\bth}_A,
\label{act-1}
\end{multline}
where
\[
E^A:={d}(\bth^A_\perp)+\vec{N}\lrcorner{d}\un{\bth}^A.
\]

For the sake of further convenience we will change the way we parameterize the configuration space. The space consists of all global coframes $\bth^A$ on $\cal M$. On the other hand the foliation ${\cal M}=\R\times\Sigma$ provides the decomposition \eqref{corep-dec}. Note now that in the action \eqref{act-1} there is no Lie derivative of $\bth^A_\perp$ with respect to $\partial_t$, which means that from the point of view of canonical formalism $\bth^A_\perp$ is not a dynamical variable, but rather a Lagrange multiplier which defines four constraints on the phase space. Since we would like to obtain an ADM-like Hamiltonian formulation of the theory we will parameterize the configuration space by $\un{\bth}^A$, the lapse $N$ and the shift $\vec{N}$ and treat $\bth^A_\perp$ as the function  \eqref{theta-0-xi-a} of these variables. Consequently, we will treat the action \eqref{act-1} as a functional depending on $\un{\bth}^A$, $N$ and $\vec{N}$, i.e., as $S[\un{\bth}^A,N,\vec{N}]$.  

\section{Hamiltonian formulation of the model}

Since the theory under consideration is formulated in terms of differential forms it will be convenient to use a Hamiltonian formalism which is adapted to forms. An outline of such a formalism based on that of \cite{ham-diff,mielke} is presented in Appendix \ref{can-forms}.

It will also be convenient to simplify the notation---since now we will denote the ``spatial'' part of the one-form $\bth^A$ by $\theta^A$, i.e,
\[
\un{\bth}^A\equiv\theta^A
\]  
and its Lie derivative with respect to $\partial_t$ by $\dot{\theta}^A$, i.e.,
\[
\LL_{\partial_t}\un{\bth}^A \equiv \dot{\theta}^A.
\]  

At the end of the previous section we reparameterized the configuration space by $\theta^A$, the lapse $N$ and the shift $\vec{N}$. Since in the action \eqref{act-1} there is no Lie derivative of $N$ and $\vec{N}$ with respect to $\partial_t$ we will treat them as Lagrange multipliers. Thus a point in the phase space of the theory consists of
\begin{enumerate}
\item a quadruplet of one-forms $(\theta^{A})$ on $\Sigma$ such that at each point $x\in\Sigma$ the rank of the matrix $(\theta^A_i(x))$ is maximal;
\item momentum $(p_A)$ conjugate to $\theta^A$: since $\Sigma$ is three dimensional and $\theta^A$ is a one-form $(p_A)$ is a quadruplet of two-forms.
\end{enumerate}         
Equivalently, a point in the phase space of the theory consists of
\begin{enumerate}
\item a map $\theta:T\Sigma\to \mathbb{M}$ such that for every $x\in\Sigma$ the restriction of $\theta$ to $T_x\Sigma$ is a linear injection;
\item the momentum $p$ as a two-form on $\Sigma$ valued in $\mathbb{M}^*$ being the dual space to $\mathbb{M}$.
\end{enumerate}
The Poisson bracket between two function $F$ and $G$ on the phase space is given by the following formula
\[
\{F,G\}=\int_\Sigma\Big(\frac{\delta F}{\delta {\theta}^A}\we\frac{\delta G}{\delta p_A}-\frac{\delta G}{\delta {\theta}^A}\we\frac{\delta F}{\delta p_A}\Big).
\]

Let us now perform the Legendre transformation and derive the Hamiltonian. Denoting by $L$ the integrand in \eqref{act-1} we define the Hamiltonian as
\[
H({\theta}^A,{\dot{\theta}}^A,N,\vec{N}):=\int_{\Sigma}{\dot{\theta}}^A\we p_A-L_\perp.
\] 
where 
\[
p_A:=\frac{\partial L_\perp}{\partial {\dot{\theta}}^A}.
\]
Direct calculation gives us
\begin{equation}
p_A=\frac{1}{N}{*}({\dot{\theta}}_A-E_A),
\label{mom}
\end{equation}
hence
\begin{multline*}
H({\theta}^A,{\dot{\theta}}^A,N,\vec{N})=\int_{\Sigma}\frac{1}{2N}({\dot{\theta}}^A+E^A)\we{*}({\dot{\theta}}_A-E_A)+\frac{N}{2}{d}{\theta}^A\we{*}{d}{\theta}_A=\\=\int_{\Sigma}\frac{1}{2N}{\dot{\theta}}^A\we{*}{\dot{\theta}}_A-\frac{1}{2N}E^A\we{*}E_A+\frac{N}{2}{d}{\theta}^A\we{*}{d}{\theta}_A.
\end{multline*}
Reversing the formula \eqref{mom} we get
\[
{\dot{\theta}}^A={N}{*}p^A+E^A,
\]
and therefore
\[
\frac{1}{2N}{\dot{\theta}}^A\we{*}{\dot{\theta}}_A=\frac{N}{2}p_A\we{*}p^A+E^A\we p_A+\frac{1}{2N}E^A\we{*}E_A.
\]
Thus
\[
H({\theta}^A,p_A,N,\vec{N})=\int_\Sigma\frac{N}{2}p^A\we{*}p_A+E^A\we p_A+\frac{N}{2}{d}{\theta}^A\we{*}{d}{\theta}_A.
\] 
On the other hand using \eqref{theta-0-xi-a} we express the one form $E^A$ as 
\[
E^A=d(N\xi^A+\vec{N}\lr\theta^A)+\vec{N}\lr d\theta^A=d(N\xi^A)+{\LL}_{\vec{N}}\theta^A,
\]
where $\LL_{\vec{N}}$ is the Lie derivative on $\Sigma$ with respect to the vector field $\vec{N}$. Consequently,
\[
E^A\we p_A=-N\xi^Adp_A+d(N\xi^Ap_A)+{\LL}_{\vec{N}}\theta^A\we p_A
\]  
and
\begin{equation*}
H({\theta}^A,p_A,N,\vec{N})=\int_\Sigma N\Big(\frac{1}{2}p^A\we{*}p_A-\xi^Adp_A+\frac{1}{2}{d}{\theta}^A\we{*}{d}{\theta}_A\Big)+({\cal L}_{\vec{N}}{\theta}^A)\we p_A,
\end{equation*}
where $\xi^A$ is a function of the canonical variable $\theta^A$ given by \eqref{xi-a} and $N$ and $\vec{N}$ are Lagrange multipliers. The last term of the Hamiltonian can be expressed as
\begin{multline}
({\cal L}_{\vec{N}}{\theta}^A)\we p_A=-{d}\,{\theta}^A\we(\vec{N}\lrcorner p_A)-(\vec{N}\lrcorner{\theta}^A)\we {d}p_A+{d}((\vec{N}_\lrcorner{\theta}^A)\we p_A)=\\=-{\theta}^A\we{\cal L}_{\vec{N}}p_A+{d}(\vec{N}\lrcorner({\theta}^A\we p_A)),
\label{L-theta-p}
\end{multline}
hence
\begin{multline}
H[{\theta}^A,p_A,N,\vec{N}]=\int_\Sigma N\Big(\frac{1}{2}p^A\we{*}p_A-\xi^Adp_A+\frac{1}{2}{d}{\theta}^A\we{*}{d}{\theta}_A\Big)-\\-{d}{\theta}^A\we(\vec{N}\lrcorner p_A)-(\vec{N}\lrcorner{\theta}^A)\we {d}p_A.
\label{ham}
\end{multline}

\section{Algebra of constraints}

The Hamiltonian \eqref{ham} depends on the Lagrange multipliers $N$ and $\vec{N}$.  Variation of the Hamiltonian with respect to the multipliers give us the following constraints:
\begin{align}
  &\frac{\delta H}{\delta N}=\frac{1}{2}p^A\we{*}p_A-\xi^Adp_A+\frac{1}{2}d{\theta}^A\we{*}{d}{\theta}_A=0,\label{sc-constr}\\ 
&\frac{\delta H}{\delta N^i}=-{d}{\theta}^A\we(\partial_i\lrcorner p_A)-(\partial_i\lrcorner{\theta}^A)\we {d}p_A=0.\label{vc-constr}
\end{align}
The constraints can be equivalently expressed as functionals on the phase space---for every function $M$ on $\Sigma$ and for every vector field $\vec{M}$ on $\Sigma$ 
\begin{align*}
&S(M):=\int_\Sigma M\frac{\delta H}{\delta N}=\int_\Sigma M\Big(\frac{1}{2}p^A\we{*}p_A-\xi^Adp_A+\frac{1}{2}d{\theta}^A\we{*}{d}{\theta}_A\Big)=0,\\
&V(\vec{M}):=\int_\Sigma M^i\frac{\delta H}{\delta N^i}=\int_\Sigma-{d}{\theta}^A\we(\vec{M}\lrcorner p_A)-(\vec{M}\lrcorner{\theta}^A)\we {d}p_A=0.
\end{align*}
We will call $S(M)$ a {\em scalar} constraint and $V(\vec{M})$ a {\em vector} constraint. 
Now the Hamiltonian \eqref{ham} can be written as
\begin{equation}
H[{\theta}^A,p_A,N,\vec{N}]=S(N)+V(\vec{N}).
\label{ham-sv}
\end{equation}

The goal of this section is to show that $(i)$ the vector and the scalar constraints are the only constraints of the system and $(ii)$ they are of the first class. To reach the goals we have to calculate Poisson brackets between the constraints. 

To make the calculations easier and more transparent we are going to introduce some auxiliary formulae. 

\subsection{Auxiliary formulae}

Let $\alpha$ be a one-form on $\Sigma$. The vector field obtained from $\alpha$ by raising its index with the inverse metric \eqref{q-1} will be denoted by $\vec{\alpha}$:
\[
\vec{\alpha}=\alpha^i\partial_i:=\alpha_iq^{ij}\partial_j.
\]    

Let $\beta$ be a $k$-form on $\Sigma$ and $\alpha$ a one-form on the manifold. Then, as proven in Appendix \ref{Hodge-app},
\begin{equation}
*(*\beta\we\alpha)=\vec{\alpha}\lr\beta.
\label{a-b}
\end{equation}   

The next important formula is one describing a functional derivative of the Hodge operator $*$. More precisely, assume that $\alpha$ and $\beta$ are $k$-forms on $\Sigma$ independent of the canonical variables $\theta^A$ and $p_B$. Then
\begin{equation}
\frac{\delta}{\delta{\theta}^A}\int_\Sigma\alpha\we*\beta=\vec{\theta}^B\lrcorner\Big(\eta_{AB}\,\alpha\we{*}\beta-(\vec{\theta}_A\lrcorner\alpha)\we{*}(\vec{\theta}_B\lrcorner\beta)-(\vec{\theta}_B\lrcorner\alpha)\we{*}(\vec{\theta}_A\lrcorner\beta)\Big).
\label{a*b-fder}
\end{equation}      
For the proof of this equation see Appendix \ref{f-der-*}. Taking into account the complexity of the r.h.s. of the equation it will be convenient to introduce a short notation for it:
\begin{equation}
\vec{\theta}^B\lrcorner\Big(\eta_{AB}\,\alpha\we{*}\beta-(\vec{\theta}_A\lrcorner\alpha)\we{*}(\vec{\theta}_B\lrcorner\beta)-(\vec{\theta}_B\lrcorner\alpha)\we{*}(\vec{\theta}_A\lrcorner\beta)\Big)\equiv \alpha \we\Star{A}\beta.
\label{a*'b}
\end{equation}
Let us emphasize that the symbol $\alpha \we\Star{A}\beta$ as an abbreviation of the l.h.s. of \eqref{a*'b} will also be used in cases when the forms $\alpha$ and $\beta$ do depend on the canonical variables.

While calculating the Poisson brackets we will encounter a contraction of $\xi^A$ with $\alpha\we\Star{A}\beta$. It is shown in Appendix \ref{f-der-*} that
\begin{equation}
\xi^A(\alpha\we\Star{A}\beta)=0.
\label{xi-d*=0}
\end{equation}
   
The next formula describes the Lie derivative on $\Sigma$ of a three-form built from $k$-forms $\alpha,\beta$ and the Hodge operator. For a vector field $\vec{M}$ on $\Sigma$ we have 
\begin{equation}
\LL_{\vec{M}}(\alpha\we*\beta)=\LL_{\vec{M}}\alpha\we*\beta+\alpha\we*\LL_{\vec{M}}\beta+\LL_{\vec{M}}\theta^A\we(\alpha\we\Star{A}\beta).
\label{L-*}
\end{equation}
A proof of this equation can be found in Appendix \ref{L-*-app}. 
 
The last formula,
\begin{equation}
\frac{1}{2}\veps^D{}_{BCA}\theta^B\we\theta^C \xi^A=-*\theta^D,
\label{ett-xi}
\end{equation}
is proven in Appendix \ref{Hodge-app}.

\subsection{Poisson bracket of vector constraints}

It follows from \eqref{L-theta-p} that 
\[
V(\vec{M})=\int_\Sigma p_A\we({\cal L}_{\vec{M}}{\theta}^A)=-\int_\Sigma{\theta}^A\we{\cal L}_{\vec{M}}p_A,
\]
hence
\[
\frac{\delta V(\vec{M})}{\delta p_A}={\cal L}_{\vec{M}}{\theta}^A, \ \ \frac{\delta V(\vec{M})}{\delta {\theta}^A}=-{\cal L}_{\vec{M}}p_A.
\]
Thus
\begin{multline*}
\{V(\vec{M}),V(\vec{M}')\}=\int_\Sigma-\LL_{\vec{M}}p_A\we\LL_{\vec{M}'}\theta^A-(\vec{M}\leftrightarrow \vec{M}')=\\=\int_\Sigma-\LL_{\vec{M}}(p_A\we\LL_{\vec{M}'}\theta^A)+p_A\we\LL_{\vec{M}}\LL_{\vec{M}'}\theta^A-(\vec{M}\leftrightarrow \vec{M}')=\\=\int_{\Sigma} p_A\we[\LL_{\vec{M}},\LL_{\vec{M}'}]\theta^A=\int_{\Sigma} p_A\we\LL_{[\vec{M},\vec{M}']}\theta^A=V([\vec{M},\vec{M}'])
\end{multline*}
---here we used the fact that 
\begin{equation}
\int\LL_{\vec{M}}\alpha=0
\label{L-a=0}
\end{equation}
for every three-form $\alpha$ on $\Sigma$.

\subsection{Poisson bracket of scalar constraints}

Calculation of the Poisson bracket $\{S(M),S(M')\}$ is more difficult. Let us introduce the following three functionals
\begin{align*}
&S_1(M):=\int_\Sigma \frac{M}{2}p^A\we{*}p_A,\\
&S_2(M):=-\int_\Sigma M\xi^A{d}p_A,\\
&S_3(M):=\int_\Sigma \frac{M}{2}{d}{\theta}^A\we{*}{d}{\theta}_A.
\end{align*} 
Then
\[
S(M)=S_1(M)+S_2(M)+S_3(M)
\]
and
\begin{multline}
\{S(M),S(M')\}=\{S_1(M),S_1({M'})\}+\{S_2(M),S_2({M'})\}+\{S_3(M),S_3({M'})\}+\\\Big(\{S_1({M}),S_2({M'})\}+\{S_2({M}),S_3({M'})\}+\{S_3({M}),S_1({M'})\}-(M\leftrightarrow M')\Big).
\label{SMSM}
\end{multline}

The functional derivatives of the three functionals are of the following form
\begin{align*}
&\frac{\delta S_1(M)}{\delta {\theta}^A}=\frac{M}{2}p_B\we\Star{A}p^B,\\
&\frac{\delta S_1(M)}{\delta p_A}={M}{*}p^A,\\
&\frac{\delta S_2(M)}{\delta {\theta}^A}=\frac{M}{2}({*}{d}p_{D})\veps^{D}{}_{BCA}{\theta}^{B}\we{\theta}^{C}+\frac{M}{3!}\veps^{B}{}_{CDE}\,[dp_B\we\Star{A}({\theta}^{C}\we{\theta}^{D}\we{\theta}^{E})],\\
&\frac{\delta S_2(M)}{\delta p_A}={d}(M\xi^A),\\
&\frac{\delta S_3(M)}{\delta {\theta}^A}={d}(M{*}{d}{\theta}_A)+\frac{M}{2}d\theta^B\we\Star{A}d\theta_B,\\
&\frac{\delta S_3(M)}{\delta p_A}=0.
\end{align*}

Although the above derivatives appear to be quite complicated functions of the canonical variables it is not very difficult to see that most terms in \eqref{SMSM} vanish. Indeed, the ``quadratic'' bracket $\{S_3(M),S_3(M')\}$ vanishes because both functionals do not depend on the momentum $p_A$. Another ``quadratic'' one
\[
\{S_1(M),S_1({M'})\}=\int_\Sigma\frac{MM'}{2}(p_B\we\Star{A}p^B)\we{*}p^A-(M\leftrightarrow M')=0.
\]

Next we consider ``mixed'' terms $\{S_1(M),S_2(M')\}$ and $\{S_2(M),S_3(M')\}$:  
\begin{multline*}
\{S_1(M),S_2(M')\}=\int_\Sigma\frac{M}{2}(p_B\we\Star{A}p^B)\we(\xi^AdM'+M'd\xi^A)-\frac{\delta S_2(M)}{\delta {\theta}^A}\we\frac{\delta S_1(M')}{\delta p_A}=\\=\int_\Sigma\frac{MM'}{2}(p_B\we\Star{A}p^B)\we d\xi^A-\frac{\delta S_2(M)}{\delta {\theta}^A}\we\frac{\delta S_1(M')}{\delta p_A},
\end{multline*}
where in the last step we used \eqref{xi-d*=0}. The two obtained terms are proportional to $MM'$ hence
\[
\{S_1(M),S_2(M')\}-(M\leftrightarrow M')=0.
\]
The other ``mixed'' term
\begin{multline*}
\{S_2(M),S_3(M')\}=-\int_\Sigma\Big({d}(M'{*}{d}{\theta}_A)+\frac{M'}{2}(d\theta^B\we\Star{A}d\theta_B)\Big)\we {d}(M\xi^A)=\\=-\int_\Sigma\frac{M'}{2}(d\theta^B\we\Star{A}d\theta_B)\we(\xi^AdM+Md\xi^A)=-\int_\Sigma\frac{M'M}{2}(d\theta^B\we\Star{A}d\theta_B)\we d\xi^A,
\end{multline*}
where in the second step we used the Stokes theorem and in the last one we applied \eqref{xi-d*=0}. Thus
\[
\{S_2(M),S_3(M')\}-(M\leftrightarrow M')=0.
\]

It turns out that the remaining two terms in \eqref{SMSM} do not vanish. Let us begin with the ``quadratic'' term
\begin{multline*}
\{S_2(M),S_2({M'})\}=\int_\Sigma \Big(\frac{M}{2}({*}{d}p_{D})\veps^{D}{}_{BCA}{\theta}^{B}\we{\theta}^{C}+\\+\frac{M}{3!}\veps^{B}{}_{CDE}\,\Big(dp_B\we\Star{A}({\theta}^{C}\we{\theta}^{D}\we{\theta}^{E})\Big)\Big)\we(\xi^AdM'+M'd\xi^A)-\\-(M\leftrightarrow M')=\int_\Sigma\frac{1}{2}({*}{d}p_{D})\veps^{D}{}_{BCA}\xi^A{\theta}^{B}\we{\theta}^{C}\we m=-\int_\Sigma (*dp_D) *\theta^D\we m=\\=-\int_\Sigma *(*\theta^D\we m)\we dp_D=-\int_\Sigma \vec{m}\lr \theta^A\we dp_A. 
\end{multline*}
Here we used: Equation \eqref{xi-d*=0} in the second step, \eqref{ett-xi} in the third step, \eqref{a-b} in the last one and denoted
\[
m:=M{d}M'-M'{d}M.
\]
The other non-vanishing term is a ``mixed'' one:
\begin{multline*}
  \{S_3({M}),S_1({M'})\}-(M\leftrightarrow M')=\int_\Sigma \Big({d}(M{*}{d}{\theta}_A)+\frac{M}{2}d\theta^B\we\Star{A}d\theta_B\Big)\we M'*p^A-\\-(M\leftrightarrow M')=\int_\Sigma m\we{*}p^A\we{*}{d}{\theta}_A=-\int_\Sigma  d\theta^A\we*(*p_A\we m)=-\int_\Sigma d\theta^A\we\vec{m}\lr p_A,
\end{multline*}
where in the last step we used \eqref{a-b}.

Collecting the two nonzero results we get
\[
\{S(M),S({M'})\}=\int_\Sigma -d\theta^A\we\vec{m}\lr p_A -\vec{m}\lr \theta^A\we dp_A =V(\vec{m}).
\]

\subsection{Poisson bracket of vector and scalar constraints}

Clearly,
\begin{equation*}
\{S(M),V(\vec{M})\}=\sum_{i=1}^3\{S_i(M),V(\vec{M})\}.
\end{equation*}
In fact each of the three terms can be calculated in a similar way, therefore we present detailed calculation regarding only one of them:
\begin{equation*}
\{S_3(M),V(\vec{M})\}=\int_\Sigma{d}(M{*}{d}{\theta}_A)\we\LL_{\vec{M}}\theta^A+\frac{M}{2}(d\theta^B\we\Star{A}d\theta_B)\we\LL_{\vec{M}}\theta^A.
\end{equation*}
The first term can be transformed as follows
\begin{multline*}
\int_\Sigma{d}(M{*}{d}{\theta}_A)\we\LL_{\vec{M}}\theta^A=\int_\Sigma M{*}{d}{\theta}_A\we d\LL_{\vec{M}}\theta^A=\\=\int_\Sigma M{*}{d}{\theta}_A\we \LL_{\vec{M}}d\theta^A=\int_\Sigma\frac{M}{2}(\LL_{\vec{M}}d\theta^A\we{*}{d}{\theta}_A+{d}{\theta}_A\we *\LL_{\vec{M}}d\theta^A).
\end{multline*}
Thus
\begin{multline*}
\{S_3(M),V(\vec{M})\}=\int_\Sigma \frac{M}{2}(\LL_{\vec{M}}d\theta^A\we{*}{d}{\theta}_A+{d}{\theta}_A\we *\LL_{\vec{M}}d\theta^A+\LL_{\vec{M}}\theta^A\we(d\theta^B\we\Star{A}d\theta_B))=\\=\int_\Sigma \frac{M}{2}\LL_{\vec{M}}(d\theta^A\we*d\theta_A)=-\int_\Sigma\frac{1}{2}(\LL_{\vec{M}}M)d\theta^A\we*d\theta_A=-S_3(\LL_{\vec{M}}M),
\end{multline*}
where in the second step we used Equation \eqref{L-*} and in the third Equation \eqref{L-a=0}. Similarly
\[
\{S_i(M),V(\vec{M})\}=-S_i(\LL_{\vec{M}}M)
\]
for $i=1,2$. Consequently,
\[
\{S(M),V(\vec{M})\}=-S(\LL_{\vec{M}}M).
\]   

\subsection{Conclusions}

To summarize the calculations of the Poisson brackets let us collect the results:
\begin{align*}
&\{V(\vec{M}),V(\vec{M'})\}=V([\vec{M},\vec{M'}]),\\
&\{S(M),S({M'})\}=V(\vec{m}), \ \ \ m=MdM'-M'dM,\\
&\{S({M}),V({\vec{M}})\}=-S({\cal L}_{\vec{M}}M).
\end{align*}

Since the Hamiltonian \eqref{ham} is a sum of the two constraints the Poisson brackets between the constraints and the Hamiltonian vanish weakly. This means that the scalar and vector constraints are preserved by the evolution generated by the Hamiltonian hence there are no other constraints. This conclusion together with the results above means that the constraints are of {\em the first class}.  

Note finally that the vector field $\vec{m}$ appearing above depends via the inverse metric $q^{ab}$ on the configuration variable ${\theta}^A$ hence it is not a structure constant but rather a structure function. 

\section{Summary and discussion}

We showed that the theory of a cotetrad on a four-dimensional manifold $\cal M$ given by the action \eqref{act} can be easily expressed in a Hamiltonian form. A point in the phase space is a pair constituted by a restriction $(\un{\bth}^A)\equiv(\theta^A)$ of a cotetrad $(\bth^A)$ to the spatial three-dimensional manifold $\Sigma$ and a quadruplet $(p_A)$ of two-forms on $\Sigma$. The physical subset of the phase space is given by the scalar \eqref{sc-constr} and the vector \eqref{vc-constr} constraints. The Hamiltonian \eqref{ham} of the theory turned out to be a sum of the constraints which, of course, is not a surprise taking into account the fact that the action \eqref{act} is diffeomorphism invariant. The constraints are of the first class.

Let us emphasize that the Hamiltonian formulation presented in this paper is similar to the ADM formulation of GR \cite{adm} as the unphysical degrees of freedom of the initial configuration space were parameterized by the lapse $N$ and the shift $\vec{N}$ (see the remark at the end of Subsection \ref{act-dec-ss}). Consequently, the constraints appearing in the Hamiltonian formulation are the scalar and the vector ones and their algebra is not a Lie algebra since the Poisson bracket of the scalar constraints is the vector constraint smeared with a vector field which depends on the canonical variables. It was shown \cite{maluf} in the case of TEGR that if the unphysical degrees of freedom are parameterized by $\bth^A_\perp$ then the resulting constraint algebra is a true Lie algebra. It would be interesting to check whether in the case of the theory analyzed in this paper one can obtain a true Lie algebra of constraints in the same way. 
   
Let us finally comment on the structure of the scalar constraint. Let $\bld{\alpha}=\bld{\alpha}^A\ot v_A$ and $\bld{\beta}=\bld{\beta}^B\ot v_B$ be two-forms on $\cal M$ valued in $\mathbb{M}$. Given coframe $(\bth^A)$ on the manifold, which defines the Hodge operator $\star$, one can introduce bilinear map
\[
(\bld{\alpha},\bld{\beta})\mapsto \bld{K}(\bld{\alpha},\bld{\beta}):=\frac{1}{2}\bld{\alpha}^A\we\star\bld{\beta}_A.
\]        
Similarly, let ${\alpha}={\alpha}^A\ot v_A$ and ${\beta}={\beta}^B\ot v_B$ be two-forms on $\Sigma$ valued in $\mathbb{M}$. Given restricted coframe $(\un{\bth}^A)$ on the manifold, which defines the Hodge operator $*$, one can introduce another bilinear map
\[
({\alpha},{\beta})\mapsto {K}({\alpha},{\beta}):=\frac{1}{2}{\alpha}^A\we*{\beta}_A.
\]
Note now that the action \eqref{act} can be written as
\[
S[\bth^A]=-\int \bld{K}(\bed \bth,\bed \bth),
\]
while the scalar constraint can be expressed as 
\[
S(M)=\int {K}(p,p)-\xi^A dp_A +{K}(d\un{\bth},d\un{\bth}),
\]
where $\bed \bth=\bed \bth^A\ot v_A$ and $d\un{\bth}=d\un{\bth}^B\ot v_B$. Thus we see that two of the three terms constituting the scalar constraint are closely related to the action \eqref{act}. 
What is interesting about this is that the structure of a scalar constraint appearing in the Hamiltonian formulation of TEGR obtained in \cite{oko} is similar: the scalar constraint consists of three terms: one is $-\xi^A dp_A$, while the remaining two are related in an analogous way to the action \eqref{tegr-act} being the departure point of the analysis presented in \cite{oko}.     

\paragraph{Acknowledgments} This work was partially supported by the grants N N202 104838 and 182/N-QGG/2008/0 (PMN) of Polish Ministerstwo Nauki i Szkolnictwa Wy\.zszego.

\appendix

\section{Volume form \label{vol}}

Let $V$ be a real $n$-dimensional oriented vector space equipped with a scalar product $g$. Suppose that the signature of $g$ is $m\in\{0,1,2,\ldots,n\}$, i.e., in every basis of $V$ orthonormal with respect to $g$ there are exactly $m$ vectors normed to $-1$ (and $(n-m)$ vectors normed to $1$). If $(\omega^\mu)$ ($\mu=1,\ldots,n$) is a basis dual to an orthonormal basis of $V$ compatible with the orientation of $V$ then
\[
\eps:=\omega^1\we\ldots\we\omega^n
\]                 
is a volume form on $V$ given by $g$. 

Assuming that we use $g$ and its inverse $g^{-1}$ to, respectively, lower and raise indeces of components of tensors over $V$ then the following formula holds
\begin{equation}
\epsilon_{\rho_{1}\ldots \rho_{k}\nu_{1}\ldots \nu_{l}}\epsilon^{\rho_{1}\ldots \rho_{k}\mu_{1}\ldots \mu_{l}}=(-1)^{m}\,k!\,l!\,\delta^{[\mu_{1}}\!_{\nu_{1}}\ldots \delta^{\mu_{l}]}\!_{\nu_{l}},
\label{eps-eps!}
\end{equation}
where $k,l\in{0,1,2,\ldots,n}$ satisfy $k+l=n$. 

\begin{proof}[Proof of \eqref{eps-eps!}]
Note first that both tensors $\epsilon_{\rho_{1}\ldots \rho_{k}\nu_{1}\ldots \nu_{l}}\epsilon^{\rho_{1}\ldots \rho_{k}\mu_{1}\ldots \mu_{l}}$  and  $\delta^{[\mu_{1}}\!_{\nu_{1}}\ldots \delta^{\mu_{l}]}\!_{\nu_{l}}$ appearing in \eqref{eps-eps!}
\begin{enumerate}
\item are antisymmetric with respect to both upper indices and lower indices;
\item their components are nonzero if and only if $(i)$ the indices $\{\mu_{1},\ldots, \mu_{l}\}$ are pairwise distinct and $(ii)$ the {\em unordered} sets $\{\mu_{1},\ldots ,\mu_{l}\}$ and $\{\nu_{1},\ldots ,\nu_{l}\}$ coincide.    
\end{enumerate} 
These properties imply that there exists a function
\[
(\mu_{1},\ldots ,\mu_{l},\nu_{1},\ldots ,\nu_{l})\mapsto\lambda(\mu_{1},\ldots ,\mu_{l},\nu_{1},\ldots ,\nu_{l})
\]  
symmetric with respect to indices $\{\mu_{1},\ldots ,\mu_{l}\}$ and symmetric with respect to indices $\{\nu_{1},\ldots ,\nu_{l}\}$ such that
\[
\epsilon_{\rho_{1}\ldots \rho_{k}\nu_{1}\ldots \nu_{l}}\epsilon^{\rho_{1}\ldots \rho_{k}\mu_{1}\ldots \mu_{l}}=\lambda(\mu_{1},\ldots ,\mu_{l},\nu_{1},\ldots ,\nu_{l})\,\delta^{[\mu_{1}}\!_{\nu_{1}}\ldots \delta^{\mu_{l}]}\!_{\nu_{l}},
\] 
Suppose now that the components in the equation above are given by an orthonormal basis of $V$. Then $\eps_{1 2\ldots n}=1$ and $\eps^{1 2\ldots n}=(-1)^m$ and setting $\mu_i=\nu_i$ in the equation we obtain
\[
(-1)^mk!=\lambda(\mu_{1},\ldots ,\mu_{l},\mu_{1},\ldots ,\mu_{l})\frac{1}{l!}.
\]    
Using the symmetricity of $\lambda$ and the two properties of the tensors $\epsilon_{\rho_{1}\ldots \rho_{k}\nu_{1}\ldots \nu_{l}}\epsilon^{\rho_{1}\ldots \rho_{k}\mu_{1}\ldots \mu_{l}}$  and  $\delta^{[\mu_{1}}\!_{\nu_{1}}\ldots \delta^{\mu_{l}]}\!_{\nu_{l}}$ listed above we arrive at \eqref{eps-eps!}.
\end{proof}

\section{Hodge dualization \label{Hodge-app}}

Let $\alpha,\beta$ be $k$-forms over $V$ and let $l=n-k$, where $n=\dim V$. The scalar product $g$ defines a scalar product     
\begin{equation}
\scal{\alpha}{\beta}:=\frac{1}{k!}\alpha_{\mu_1\ldots \mu_k}g^{\mu_1\nu_1}\ldots g^{\mu_k \nu_k}\beta_{\nu_1\ldots \nu_k}=\frac{1}{k!}\alpha^{\mu_1\ldots \mu_k}\beta_{\mu_1\ldots \mu_k}.
\label{sc-pr}
\end{equation}
The Hodge operator $*$ maps a $k$-form $\beta$ to an $l$-form $*\beta$ such that for every $k$-form $\alpha$
\begin{equation}
\alpha\we *\beta=\scal{\alpha}{\beta}\eps.
\label{hodge}
\end{equation}       
Equivalently,
\begin{equation}
(*\beta)_{\nu_1\ldots\nu_l}=\frac{1}{k!}\beta_{\mu_1\ldots\mu_k}{\eps^{\mu_1\ldots\mu_k}}_{\nu_1\ldots\nu_{l}}.
\label{hodge-comp}
\end{equation}
The map $\beta\mapsto*\beta$ is a linear isomorphism between the linear space of $k$-forms and the linear space of $l$-forms satisfying
\[
**\beta=(-1)^{lk+m}\beta
\]   
for every $k$-form $\beta$.  

\begin{proof}[Proof of \eqref{A*B-dec} in a general case] Assume that ${\cal M}$ is an $n$-dimensional oriented manifold with a Lorentzian metric $g$. Suppose, moreover, that ${\cal M}=\R\times \Sigma$ and that this decomposition satisfies all the assumptions listed at the beginning of Section \ref{3+1} (modulo existence of an appropriate coframe $(\bth^A)$ which is irrelevant here).       

If $\alpha,\beta$ are $k$-forms ($k\geq 2$) on $\cal M$ then by virtue of \eqref{sc-pr}
\begin{multline}
\scal{\alpha}{\beta}=\frac{1}{k!}(k\alpha_{0i_2\ldots i_k}\beta_{0j_2\ldots j_k}g^{00}g^{i_2j_2}\ldots g^{i_kj_k}+\\+k(k-1)\alpha_{0i_2i_3\ldots i_k}\beta_{j_10j_3\ldots j_k}g^{0j_1}g^{i_20}g^{i_3j_3}\ldots g^{i_kj_k}+k\alpha_{0i_2\ldots i_k}\beta_{j_1j_2\ldots j_k}g^{0j_1}g^{i_2j_2}\ldots g^{i_kj_k}+\\+k\alpha_{i_1i_2\ldots i_k}\beta_{0j_2\ldots j_k}g^{i_10}g^{i_2j_2}\ldots g^{i_kj_k}+\alpha_{i_1\ldots i_k}\beta_{j_1\ldots j_k}g^{i_1j_1}\ldots g^{i_kj_k})
\label{a-b-scal}
\end{multline}   
Under an obvious generalization \eqref{g-1-dec} is still valid hence we have
\begin{equation}
g^{00}=-N^{-2}, \ \ g^{0i}=N^{-2}N^i, \ \ g^{ij}=q^{ij}-N^{-2}N^iN^j.
\label{g-1-comp}
\end{equation}
Using these expressions we can transform the terms at the r.h.s. of \eqref{a-b-scal} as follows: 
\begin{align*}
\text{the first term} =& \frac{1}{(k-1)!}(-N^{-2}\alpha_{0i_2\ldots i_k}\beta_{0j_2\ldots j_k}q^{i_2j_2}\ldots q^{i_kj_k}+\\&+N^{-4}(k-1)\alpha_{0i_2i_3\ldots i_k}\beta_{0j_2j_3\ldots j_k}N^{i_2}N^{j_2}q^{i_3j_3}\ldots q^{i_kj_k}),\\
\text{the second term}=& \frac{1}{(k-1)!}N^{-4}(k-1)\alpha_{0i_2i_3\ldots i_k}\beta_{j_10j_3\ldots j_k}N^{i_2}N^{j_1}q^{i_3j_3}\ldots q^{i_kj_k},\\
\text{the third term}= &\frac{1}{(k-1)!}N^{-2}\alpha_{0i_2\ldots i_k}\beta_{j_1j_2\ldots j_k}N^{j_1}q^{i_2j_2}\ldots q^{i_kj_k}=N^{-2}\scal{\alpha_\perp}{\vec{N}\lr\un{\beta}}_q,\\
\text{the fourth term}= & \frac{1}{(k-1)!}N^{-2}\alpha_{i_1i_2\ldots i_k}\beta_{0j_2\ldots j_k}N^{i_1}q^{i_2j_2}\ldots q^{i_kj_k}=N^{-2}\scal{\vec{N}\lr\un{\alpha}}{\beta_\perp}_q,\\
\text{the fifth term}=& \frac{1}{k!}(\alpha_{i_1\ldots i_k}\beta_{j_1\ldots j_k}q^{i_1j_1}\ldots q^{i_kj_k}-\\ &-N^{-2}k\alpha_{i_1i_2\ldots i_k}\beta_{j_1j_2\ldots j_k}N^{i_1}N^{j_1}q^{i_2j_2}\ldots q^{i_kj_k})=\\ = & \scal{\un{\alpha}}{\un{\beta}}_q -N^{-2}\scal{\vec{N}\lr\un{\alpha}}{\vec{N}\lr\un{\beta}}_q,
\end{align*}
where $\scal{\cdot}{\cdot}_q$ denotes a scalar product \eqref{sc-pr} of forms on $\Sigma_{t}$ given by the metric $q$ induced on the submanifold by $g$. It is easy to see that
\[
\text{the first term}+\text{the second one}=-N^{-2}\scal{\alpha_\perp}{\beta_\perp}_q.
\]     
Consequently,
\begin{multline}
\scal{\alpha}{\beta}=-N^{-2}(\scal{\alpha_\perp}{\beta_\perp}_q-\scal{\alpha_\perp}{\vec{N}\lr\un{\beta}}_q-\scal{\vec{N}\lr\un{\alpha}}{\beta_\perp}_q+\scal{\vec{N}\lr\un{\alpha}}{\vec{N}\lr\un{\beta}}_q)+\scal{\un{\alpha}}{\un{\beta}}_q=\\=-N^{-2}\scal{\alpha_\perp-\vec{N}\lr\un{\alpha}}{\beta_\perp-\vec{N}\lr\un{\beta}}_q+\scal{\un{\alpha}}{\un{\beta}}_q.
\label{scal-q}
\end{multline}

For one-forms $\alpha,\beta$ we have
\[
\scal{\alpha}{\beta}=\alpha_0\beta_0g^{00}+\alpha_0\beta_ig^{0i}+\alpha_i\beta_0g^{i0}+\alpha_i\beta_jg^{ij}.
\]
Applying \eqref{g-1-comp} one easily arrives at \eqref{scal-q}. 

If $\alpha$ is a zero-form then $\alpha=\un{\alpha}$ and $\alpha_\perp=0=\vec{N}\lr\un{\alpha}$. Therefore, for zero-forms $\alpha,\beta$ 
\[
\scal{\alpha}{\beta}=\alpha\beta=\un{\alpha}\un{\beta}=\scal{\un{\alpha}}{\un{\beta}}_q 
\]
and this equation coincides with \eqref{scal-q}.
 
Under an obvious generalization the decomposition \eqref{ee-dec} still holds in the general case. Using \eqref{ee-dec} and \eqref{scal-q} in the following equation
\begin{equation*}
\alpha\we\star\beta=\scal{\alpha}{\beta}\bld{\eps}
\end{equation*}
we obtain \eqref{A*B-dec}.
\end{proof}

\begin{proof}[Proof of \eqref{a-b}] For the sake of generality let us assume that $\Sigma$ is an $n$-dimensional pseudo-Riemannian manifold  with a metric $q$ of signature $m$. If $\beta$ is a $k$-form on $\Sigma$ ($k>0$), $\alpha$ a one-form on the manifold and $l=n-k$ then
\begin{align*}
&*\beta=\frac{1}{k!}\beta_{a_1\ldots a_k}\eps^{a_1\ldots a_k}{}_{b_1\ldots b_l}dx^{b_1}\ot\ldots\ot dx^{b_l}=\frac{1}{l!k!}\beta_{a_1\ldots a_k}\eps^{a_1\ldots a_k}{}_{b_1\ldots b_l}dx^{b_1}\we\ldots\we dx^{b_l},\\
&*\beta\we\alpha=\frac{1}{l!k!}\beta_{a_1\ldots a_k}\eps^{a_1\ldots a_k}{}_{b_1\ldots b_l}\alpha_adx^{b_1}\we\ldots\we dx^{b_l}\we dx^a=\\&=\frac{(l+1)!}{l!k!}\beta_{a_1\ldots a_k}\eps^{a_1\ldots a_k}{}_{[b_1\ldots b_l}\alpha_{a]}dx^{b_1}\ot\ldots\ot dx^{b_l}\ot dx^a,\\
&*(*\beta\we\alpha)=\frac{1}{l!k!}\beta_{a_1\ldots a_k}\eps^{a_1\ldots a_k}{}_{b_1\ldots b_l}\alpha_{a}\eps^{b_1\ldots b_la}{}_{c_1\ldots c_{k-1}}dx^{c_1}\ot\ldots\ot dx^{c_{k-1}}=\\&=\frac{1}{(k-1)!l!k!}\alpha^{a}\beta_{a_1\ldots a_k}\eps^{a_1\ldots a_k}{}_{b_1\ldots b_l}\eps^{b_1\ldots b_l}{}_{ac_1\ldots c_{k-1}}dx^{c_1}\we\ldots\we dx^{c_{k-1}}
\end{align*}
Applying \eqref{eps-eps!} we obtain
\begin{multline*}
*(*\beta\we\alpha)=\frac{(-1)^{m+kl}}{(k-1)!}\alpha^{a}\beta_{a_1\ldots a_k}\delta^{[a_1}{}_a\delta^{a_2}{}_{c_1}\ldots\delta^{a_k]}{}_{c_{k-1}}dx^{c_1}\we\ldots\we dx^{c_{k-1}}=\\=\frac{(-1)^{m+kl}}{(k-1)!}\alpha^{a_1}\beta_{a_1\ldots a_k}dx^{a_2}\we\ldots\we dx^{a_{k}}=(-1)^{m+kl}\vec{\alpha}\lr\beta.
\end{multline*}
For a three-dimensional Riemannian manifold the product $kl$ is always even, $m=0$ and \eqref{a-b} follows.
 
Equation \eqref{a-b} is true also in the case of a zero-form $\beta$---then its both sides are zero. 
\end{proof}

\begin{proof}[Proof of \eqref{ett-xi}] We prove the formula by a direct calculation:
\begin{multline*}
\frac{1}{2}\veps^D{}_{BCA}\theta^B\we\theta^C \xi^A=-\frac{1}{2\cdot3!}\veps^D{}_{BCA}\veps^A{}_{KIJ}*(\theta^K\we\theta^I\we\theta^J)\theta^B\we\theta^C=\\=-\frac{1}{2}*(\theta^D\we\theta_B\we\theta_C)\theta^B\we\theta^C=-\frac{1}{2}\eps^{ijk}\theta^D_{i}\theta_{Bj}\theta_{Ck}\theta^B_l\theta^C_ndx^l\we dx^n=-\frac{1}{2}\theta^D_i\eps^i{}_{ln}dx^l\we dx^n=\\=-*\theta^D,
\end{multline*}
where in the second step we used \eqref{eps-eps!} and in the fourth one we applied \eqref{q}.
\end{proof}

\section{Canonical formalism in terms of differential forms \label{can-forms}}

The formalism we are going to describe here is based on the one presented in \cite{ham-diff,mielke}.

\subsection{Variational calculus}

Denote by $\Omega^k(U)$ a space of $k$-forms on an open subset $U$ of an $n$-dimensional manifold $\Sigma$. Consider a functional 
\[
\Omega^k(U)\ni\beta\mapsto F[\beta]\in\R.
\]    
Let $l=n-k$. The functional derivative $\delta F/\delta\beta$ is a map from $\Omega^k(U)$ to $\Omega^{l}(U)$ such that for every $\delta\beta\in\Omega^k(U)$ vanishing on the boundary $\partial U$ and for every $\beta\in\Omega^k(U)$ 
\begin{equation}
(\delta F)[\beta]=\int_U\delta\beta\we\frac{\delta F}{\delta\beta}[\beta].
\label{fder-form}
\end{equation}
In the standard formalism the functional derivative  ${\delta \tilde{F}}/\delta \beta_{j_1\ldots j_k}(x)$ is a tensor density of weight $1$ such that
\begin{equation}
\delta F=\int_U\frac{\delta \tilde{F}}{\delta \beta_{a_1\ldots a_k}(x)}\delta\beta_{a_1\ldots a_k}(x)\,dx^n.
\label{fder-st}
\end{equation}
To find the relation between $\delta F/\delta\beta$ and ${\delta \tilde{F}}/\delta \beta_{j_1\ldots j_k}(x)$ let us express the r.h.s. of \eqref{fder-form} by means of a coordinate frame $(x^a)$ on $U$: 
\begin{multline*}
\int_U\delta\beta\we\frac{\delta F}{\delta\beta}[\beta]=\int_U\frac{1}{k!}\delta\beta_{a_1\ldots a_k}dx^{a_1}\we\ldots\we dx^{a_k}\we\frac{1}{l!}\Big(\frac{\delta F}{\delta\beta}\Big){}_{b_1\ldots b_l}dx^{b_1}\we\ldots \we dx^{b_l}=\\=\int_U\frac{1}{l!k!}\delta\beta_{a_1\ldots a_k}\Big(\frac{\delta F}{\delta\beta}\Big){}_{b_1\ldots b_l}\tilde{\eps}^{a_1\ldots a_kb_1\ldots b_l}\,dx^n,
\end{multline*}
where $\tilde{\eps}^{a_1\ldots a_kb_1\ldots b_l}$ is the Levi-Civita density of weight $1$ on $U$. Comparing the expression above with \eqref{fder-st} we obtain
\begin{equation}
\frac{\delta \tilde{F}}{\delta \beta_{a_1\ldots a_k}(x)}=\frac{1}{l!k!}\Big(\frac{\delta F}{\delta\beta}\Big){}_{b_1\ldots b_l}(x)\tl{\eps}^{a_1\ldots a_kb_1\ldots b_l}.
\label{st-form}
\end{equation}
To inverse the relation we use the Levi-Civita density $\,^{\text{\tiny{--1}}}\!\tl{\eps}_{a_1\ldots a_kb_1\ldots b_l}$ of weight $-1$ and the identity  
\begin{equation}
\,^{\text{\tiny{--1}}}\!\tl{\eps}_{b_1\ldots b_k a_1\ldots a_l}\tilde{\eps}^{b_{1}\ldots b_{k}c_{1}\ldots c_{l}}=k!\,l!\,\delta^{[c_{1}}\!_{a_{1}}\ldots \delta^{c_{l}]}\!_{a_{l}}
\label{tl-eps-eps}
\end{equation}
which can be easily deduced from \eqref{eps-eps!} and obtain
\begin{equation}
\frac{\delta F}{\delta\beta}=\frac{1}{l!}\frac{\delta \tilde{F}}{\delta \beta_{a_1\ldots a_k}(x)}\,^{\text{\tiny{--1}}}\!\tl{\eps}_{a_1\ldots a_kb_1\ldots b_l}\,dx^{b_1}\we\ldots\we dx^{b_l}.
\label{form-st}
\end{equation}
The formulae \eqref{st-form} and \eqref{form-st} allow us to pass from the canonical formalism in terms of differential forms to the standard one and vice versa.

\subsection{Differential calculus}

Suppose that $\gamma$ is an $n$-form ($n=\dim\Sigma$) on $\Sigma$ which depends on a $k$-form $\beta$, but is independent of $d\beta$. One can define a partial derivative $\partial\gamma/\partial\beta$ of $\gamma$ with respect to $\beta$  as an $l$-form such that
\begin{equation}
\delta\gamma=\delta\beta\we\frac{\partial \gamma}{\partial \beta}.
\label{pg-pb-df}
\end{equation}
To find a convenient expression for the derivative let us first introduce a density of weight $1$ 
\begin{equation}
\tl{\gamma}:=\frac{1}{n!}\gamma_{i_1\ldots i_n}\tl{\eps}^{i_1\ldots i_n}.
\label{tl-g}
\end{equation}
Then 
\begin{equation}
\gamma=\tl{\gamma}dx^1\we\ldots \we dx^n
\label{g-tg}
\end{equation}
and
\begin{multline}
\delta\gamma=\delta\tl{\gamma}dx^1\we\ldots \we dx^n=\frac{\partial\tl{\gamma}}{\partial \beta_{a_1\ldots a_k}}\delta\beta_{a_1\ldots a_k}dx^1\we\ldots \we dx^n=\\=\delta\beta_{b_1\ldots b_k}\frac{\partial\tl{\gamma}}{\partial \beta_{a_1\ldots a_k}}\delta^{a_1}{}_{b_1}\ldots\delta^{a_k}{}_{b_k}dx^1\we\ldots \we dx^n.
\label{dg-part}
\end{multline}
Applying \eqref{tl-eps-eps} we obtain the following formula
\begin{multline*}
\delta\gamma =\frac{1}{k!}\delta\beta_{b_1\ldots b_k}\frac{1}{l!}\frac{\partial\tl{\gamma}}{\partial \beta_{a_1\ldots a_k}}\,^{\text{\tiny{--1}}}\!\tl{\eps}_{a_1\ldots a_k c_1\ldots c_l}\tilde{\eps}^{b_{1}\ldots b_{k}c_{1}\ldots c_{l}}dx^1\we\ldots \we dx^n=\\=\frac{1}{k!}\delta\beta_{b_1\ldots b_k}\frac{1}{l!}\frac{\partial\tl{\gamma}}{\partial \beta_{a_1\ldots a_k}}\,^{\text{\tiny{--1}}}\!\tl{\eps}_{a_1\ldots a_k c_1\ldots c_l}dx^{b_1}\we\ldots \we dx^{b_k}\we dx^{c_1}\we\ldots\we dx^{c_l},
\end{multline*}
hence
\begin{equation}
\frac{\partial \gamma}{\partial \beta}=\frac{1}{l!}\frac{\partial\tl{\gamma}}{\partial \beta_{a_1\ldots a_k}}\,^{\text{\tiny{--1}}}\!\tl{\eps}_{a_1\ldots a_k c_1\ldots c_l}dx^{c_1}\we\ldots\we dx^{c_l}.
\label{pg-pb}
\end{equation}

\subsection{Canonical formalism}

Let $\alpha$ be a $k$-form  on a manifold ${\cal M}:=\R\times\Sigma$. Consider the following action 
\[
S[\alpha]=\int_{\cal M} L(\alpha,\bed \alpha),
\]  
where $L$ is an $(n+1)$-form on $\cal M$, and $\bed$ is the exterior derivative on the manifold. Assume that $(x^a)$ are (local) coordinates on $\Sigma$ and that $(x^\mu)= (t\equiv x^0,x^a)$ are (local) coordinates on $\cal M$ compatible with the decomposition ${\cal M}=\R\times\Sigma$. It follows from the decomposition of $\bed\alpha$ in \eqref{perp-un} that in the action above there is no Lie derivative of $\alpha_\perp$ with respect to $\partial_t$ and therefore from the point of view of canonical formalism $\alpha_\perp$ can be seen as a Lagrange multiplier. Thus the only dynamical variable is $\un{\alpha}$ and the Hamiltonian is given by the standard formula 
\begin{equation}
\begin{aligned}
H(\tl{p},\un{\alpha},\alpha_\perp)&:=\int_\Sigma (\dot{\un{\alpha}}_{a_1\ldots a_k}\tl{p}^{a_1\ldots a_k}-\tl{L})dx^n,\\
\tl{p}^{a_1\ldots a_k}&:=\frac{\partial \tilde{L}}{\partial \dot{\un{\alpha}}_{a_1\ldots a_k}},
\end{aligned}
\label{ham-app}
\end{equation}              
where the tensor density $\tl{p}^{a_1\ldots a_k}$ is the momentum conjugate to $\un{\alpha}$, $\dot{\un{\alpha}}$ denotes the Lie derivative of $\un{\alpha}$ with respect to $\partial_t$ and $\tl{L}$ is defined according to  \eqref{tl-g} by the Levi-Civita density $\tl{\eps}^{\mu_1\ldots\mu_{n+1}}$ on $\cal M$. 

Note that
\begin{equation}
\tl{L}=\frac{1}{(n+1)!}L_{\mu_1\ldots\mu_{n+1}}\tl{\eps}^{\mu_1\ldots\mu_{n+1}}=\frac{1}{n!}L_{0i_1\ldots i_n}\tl{\eps}^{0i_1\ldots i_n}=\widetilde{L_\perp},
\label{tlL}
\end{equation}
hence by virtue of \eqref{g-tg} we have
\[
\int_\Sigma-\tl{L}dx^n=\int_\Sigma-\widetilde{L_\perp}dx^1\we\ldots\we dx^n=\int_\Sigma-L_\perp. 
\]
On the other hand by virtue of \eqref{tlL}
\[
\int_\Sigma \dot{\un{\alpha}}_{a_1\ldots a_k}\tl{p}^{a_1\ldots a_k}dx^n=\int_\Sigma \dot{\un{\alpha}}_{a_1\ldots a_k}\frac{\partial \widetilde{L_\perp}}{\partial \dot{\un{\alpha}}_{a_1\ldots a_k}}dx^1\we\ldots\we dx^n.
\]
To proceed further with this expression note that it follows from \eqref{pg-pb-df} and \eqref{dg-part} that 
\[
\delta\beta_{a_1\ldots a_k}\frac{\partial\tl{\gamma}}{\partial \beta_{a_1\ldots a_k}}dx^1\we\ldots \we dx^n=\delta\beta\we\frac{\partial \gamma}{\partial \beta}.
\]
Setting in this formula $\delta\beta=\dot{\un{\alpha}}$, $\tl{\gamma}=\widetilde{L_\perp}$ and $\beta=\dot{\un{\alpha}}$ we see that the integral above can be expressed as   
\[
\int_\Sigma \dot{\un{\alpha}}\we \frac{\partial L_\perp}{\partial \dot{\un{\alpha}}}.
\]

These results allow us to rewrite the Hamiltonian \eqref{ham-app} as
\[
H=\int_{\Sigma} \dot{\un{\alpha}}\we \frac{\partial L_\perp}{\partial \dot{\un{\alpha}}}-L_\perp
\]
and suggest introducing a momentum $l$-form $p$ ($l=n-k$):
\begin{equation}
p:=\frac{\partial L_\perp}{\partial \dot{\un{\alpha}}}=\frac{1}{l!}\frac{\partial\widetilde{L_\perp}}{\partial \dot{\un{\alpha}}_{a_1\ldots a_k}}\,^{\text{\tiny{--1}}}\!\tl{\eps}_{a_1\ldots a_k c_1\ldots c_l}dx^{c_1}\we\ldots\we dx^{c_l}.
\label{p}
\end{equation}
Using \eqref{pg-pb} and \eqref{tlL} one can easily find a relation between the components $p_{a_1\ldots a_l}$ and $\tl{p}^{b_1\ldots b_k}$:
\begin{equation}
p_{a_1\ldots a_l}=\tl{p}^{b_1\ldots b_k}\,^{\text{\tiny{--1}}}\!\tl{\eps}_{b_1\ldots b_k a_1\ldots a_l}.
\label{p-tlp}
\end{equation}
Now the Hamiltonian can be expressed as
\[
H(p,\un{\alpha},\alpha_\perp)=\int_{\Sigma}\dot{\un{\alpha}}\we p-L_\perp.
\]
 
A point in the phase space of the theory can be viewed now as a pair $(\un{\alpha},p)$, where $\un{\alpha}$ is a $k$-form on $\Sigma$, $p$ is an $l$-form on the manifold and $k+l=n=\dim\Sigma$.        

Let us finally find an expression for a Poisson bracket in terms of differential forms.  The bracket of functionals $F,G$ in the standard formalism reads
\begin{equation}
\{F,G\}=\int_\Sigma \Big(\frac{\delta \tl{F}}{\delta\un{\alpha}_{a_1\ldots a_k}(x)}\frac{\delta \tl{G}}{\delta \tl{p}^{a_1\ldots a_k}(x)}-\frac{\delta\tl{G}}{\delta\un{\alpha}_{a_1\ldots a_k}(x)}\frac{\delta\tl{F}}{\delta \tl{p}^{b_1\ldots b_k}(x)}\Big)\,dx^n.
\label{Pb}
\end{equation}
Then by virtue of \eqref{p-tlp} and \eqref{st-form}
\[
\frac{\delta \tl{G}}{\delta \tl{p}^{a_1\ldots a_k}(x)}=\frac{\delta \tl{G}}{\delta p_{b_1\ldots b_l}(x)}\,^{\text{\tiny{--1}}}\!\tl{\eps}_{a_1\ldots a_k b_1\ldots b_l}=(-1)^{kl}\Big(\frac{\delta G}{\delta p}\Big){}_{a_1\ldots a_k}(x),
\]
hence 
\begin{multline*}
\int_\Sigma \frac{\delta \tl{F}}{\delta\un{\alpha}_{a_1\ldots a_k}(x)}\frac{\delta \tl{G}}{\delta \tl{p}^{a_1\ldots a_k}(x)}\,dx^n=\\=\int_\Sigma \frac{(-1)^{kl}}{l!k!}\frac{\delta F}{\delta\un{\alpha}}{}_{b_1\ldots b_l}(x)\tl{\eps}^{a_1\ldots a_kb_1\ldots b_l}\Big(\frac{\delta G}{\delta p}\Big){}_{a_1\ldots a_k}(x)dx^1\we\ldots \we dx^n=\int\frac{\delta F}{\delta\un{\alpha}}\we\frac{\delta G}{\delta p},
\end{multline*}
where in the first step we applied \eqref{st-form}. This means that
\[
\{F,G\}=\int_\Sigma \Big(\frac{\delta F}{\delta\un{\alpha}}\we\frac{\delta G}{\delta p}-\frac{\delta G}{\delta\un{\alpha}}\we\frac{\delta F}{\delta p}\Big).
\]
   
\section{Functional derivative of the Hodge operator \label{f-der-*}}

The goal of this section is to prove Equations \eqref{a*b-fder} and \eqref{xi-d*=0}. 

\begin{proof}[Proof of \eqref{a*b-fder}] 
For the sake of generality we assume that $\Sigma$ is an $n$-dimensional oriented manifold equipped with a metric $q$ of signature $m$. We suppose, moreover, that there exist $s\geq n$ one-forms $\theta^A$, $A=0,1,\ldots,s-1$, on $\Sigma$ such that
\begin{equation}
q=\eta_{AB}\theta^A\ot\theta^B, 
\label{q-ett}
\end{equation}
where $(\eta_{AB})$ is a constant symmetric invertible $s\times s$-matrix (the matrix $(\eta_{AB})$ and its inverse $(\eta^{AB})$ will be used to, respectively, lower and raise the capital letter indices). 

Let
\[
F:=\int_\Sigma\alpha\we{*}\beta,
\]
where $\alpha,\beta$ are $k$-forms on $\Sigma$ {\em independent of} the forms $(\theta^A)$  and ${*}$ is the Hodge operator defined by $q$. 

From the definition \eqref{hodge} of the Hodge operator and the following expression of the volume form $\eps$ defined by $q$
\begin{equation}
\eps=\sqrt{(-1)^m\det q_{ab}}\,dx^1\we\ldots\we dx^n
\label{eps-det-q}
\end{equation}
we have
\[
\alpha\we{*}\beta=\frac{1}{k!}\alpha_{a_1\ldots a_k}\beta_{b_1\ldots b_k}q^{a_1b_1}\ldots q^{a_kb_k}\sqrt{(-1)^m\det q_{ab}}\,dx^1\we\ldots\we dx^n.
\]
Consequently, 
\[
F=\int_\Sigma \frac{1}{k!}\alpha_{a_1\ldots a_k}\beta_{b_1\ldots b_k}q^{a_1b_1}\ldots q^{a_kb_k}\sqrt{(-1)^m\det q_{ab}}\,dx^n.
\] 
If $\alpha,\beta$ do not depend on $(\theta^A)$ and $k>0$ then 
\begin{multline*}
\frac{\delta \tl{F}}{\delta \theta^A_i(x)}=\frac{1}{(k-1)!}\alpha_{a_1\ldots a_k}\beta_{b_1\ldots b_k}\frac{\partial q^{a_1b_1}}{\partial \theta^A_i(x)}\ldots q^{a_kb_k}\sqrt{(-1)^m\det q_{ab}}+\\+\frac{1}{k!}\alpha_{a_1\ldots a_k}\beta_{b_1\ldots b_k}q^{a_1b_1}\ldots q^{a_kb_k}\frac{\partial \sqrt{(-1)^m\det q_{ab}}}{\partial \theta^A_i(x)}.
\end{multline*}

Let us now find the two derivatives appearing in the last equation. Differentiating both sides of the identity $q^{ac}q_{cb}=\delta^a{}_b$ with respect to $\theta^A_i$ gives us  
\[
\frac{\partial q^{ab}}{\partial \theta^A_i}=-q^{ac}q^{bd}\frac{\partial q_{cd}}{\partial\theta^A_i}.
\]
By virtue of \eqref{q-ett} $q_{cd}=\eta_{CD}\theta^C_c\theta^D_d$ hence
\[
\frac{\partial q^{ab}}{\partial \theta^A_i}=-q^{ac}q^{bd}\eta_{CD}(\delta^C{}_A\delta^c{}_i\theta^D_d+\theta^C_c\delta^D{}_A\delta^d{}_i)=-(q^{ia}\theta^b_A+q^{ib}\theta^a_A)=-\theta^{Bi}(\theta^a_B\theta^b_A+\theta^b_B\theta^a_A),
\]
where the last step holds due to $q^{ab}=\theta^{Aa}\theta^b_A$. On the other hand
\begin{multline}
\frac{\partial \sqrt{(-1)^m\det q_{ab}}}{\partial \theta^A_i}=\frac{(-1)^m}{2\sqrt{(-1)^m\det q_{ab}}}\frac{\partial(\det q_{ab})}{\partial q_{cd}}\frac{\partial q_{cd}}{\partial\theta^A_i}=\\=\frac{(-1)^m}{2\sqrt{(-1)^m\det q_{ab}}}(q^{cd}\det q_{ab})(2\eta_{CD}\theta^C_c\delta^D{}_A\delta^i{}_d)=\sqrt{(-1)^m\det q_{ab}}\,\theta^i_A.
\label{psdq-pt}
\end{multline}
Thus we obtain
\[
\frac{\delta \tl{F}}{\delta \theta^A_i(x)}=(-\scal{\vec{\theta}_B\lrcorner\alpha}{\vec{\theta}_A\lrcorner\beta}-\scal{\vec{\theta}_A\lrcorner\alpha}{\vec{\theta}_B\lrcorner\beta}+\scal{\alpha}{\beta}\eta_{AB})\theta^{iB}\sqrt{(-1)^m\det q_{ab}},
\]
where
\[
\vec{\theta}_A:=\theta^i_A \partial_i=q^{ij}\eta_{AB}\theta^B_j\partial_i.
\]
Applying \eqref{form-st} and \eqref{eps-det-q} we easily obtain the final result \eqref{a*b-fder} for $k>0$.

Now let us show that \eqref{a*b-fder} holds also in the case $k=0$. Then
\[
F=\int_\Sigma \alpha\beta\sqrt{(-1)^m\det q_{ab}}\,dx^n
\]
and
\[
\frac{\delta \tl{F}}{\delta \theta^A_i(x)}=\scal{\alpha}{\beta}\eta_{AB}\theta^{iB}\sqrt{(-1)^m\det q_{ab}}.
\]
This means that for $k=0$ 
\[
\frac{\delta}{\delta{\theta}^A}\int_\Sigma\alpha\we*\beta=\eta_{AB}\vec{\theta}^B\lrcorner (\alpha\we{*}\beta).
\]
But for every zero-form $\alpha$ the contraction $\vth^A\lr\alpha$ is zero and the r.h.s. of the equation above coincides with the r.h.s. of \eqref{a*b-fder}. 
\end{proof}

\begin{proof}[Proof of \eqref{xi-d*=0}] The form $\alpha\we\Star{A}\beta$, that is the l.h.s. of \eqref{a*'b} can be written as follows:
\[
\alpha\we\Star{A}\beta=\theta_{Ai}\gamma^i{}_{jk}dx^j\we dx^k,
\]
where $\gamma^i{}_{jk}$ is a tensor field depending on $\theta^B$. Consequently,
\[
\xi^A(\alpha\we\Star{A}\beta)=\xi^A\theta_{Ai}\gamma^i{}_{jk}dx^j\we dx^k=0
\]
by virtue of \eqref{xi-df}.
\end{proof}

\section{Lie derivative of $\alpha\we*\beta$ \label{L-*-app}}

\begin{proof}[Proof of \eqref{L-*}] Let $\Sigma$ be an $n$-dimensional oriented manifold equipped with a metric $q$ of signature $m$ and $s\geq n$ one-forms $(\theta^A)$ such that \eqref{q-ett} is satisfied. Let $\alpha,\beta$ be $k$-forms and $\vec{M}$ a vector fields on the manifolds. It follows from \eqref{hodge} that        
\[
\LL_{\vec{M}}(\alpha\we*\beta)=\LL_{\vec{M}}(\scal{\alpha}{\beta})\eps +\scal{\alpha}{\beta}\LL_{\vec{M}}\eps.
\]
Using \eqref{sc-pr} we get for $k>0$ 
\begin{multline}
\LL_{\vec{M}}(\scal{\alpha}{\beta})=\frac{1}{k!}(\LL_{\vec{M}}\alpha)_{a_1\ldots a_k}\beta_{b_1\ldots b_k}q^{a_1b_1}\ldots q^{a_kb_k}+\\+\frac{1}{k!}\alpha_{a_1\ldots a_k}(\LL_{\vec{M}}\beta)_{b_1\ldots b_k}q^{a_1b_1}\ldots q^{a_kb_k}+\frac{1}{(k-1)!}\alpha_{a_1\ldots a_k}\beta_{b_1\ldots b_k}(\LL_{\vec{M}}q^{-1})^{a_1b_1}\ldots q^{a_kb_k},
\label{L-sc-ab}
\end{multline}
where in order to avoid any confusion the symbol $q^{-1}$ was used to denote the metric inverse to $q$ (i.e. $(q^{-1})^{ab}\equiv q^{ab}$). Because $q^{ac}q_{cb}=\delta^a{}_b$ and the Lie derivative of $\delta^a{}_b$ is zero     
\begin{multline*}
(\LL_{\vec{M}}q^{-1})^{ab}=-q^{ac}q^{bd}(\LL_{\vec{M}}q)_{cd}=-q^{ac}q^{bd}((\LL_{\vec{M}}\theta^A)_c\theta_{Ad}+(\LL_{\vec{M}}\theta^A)_d\theta_{Ac})=\\=-(q^{ac}\theta^b_A+q^{bc}\theta^a_A)(\LL_{\vec{M}}\theta^A)_c=-\theta^{Bc}(\theta^a_B\theta^b_A+\theta^b_B\theta^a_A)(\LL_{\vec{M}}\theta^A)_c.
\end{multline*}
Thus
\[
\LL_{\vec{M}}(\scal{\alpha}{\beta})=\scal{\LL_{\vec{M}}\alpha}{\beta}+\scal{\alpha}{\LL_{\vec{M}}\beta}-\Big(\scal{\vth_A\lr\alpha}{\vth_B\lr\beta}+\scal{\vth_B\lr\alpha}{\vth_A\lr\beta}\Big)\vth^{B}\lr\LL_{\vec{M}}\theta^A
\]
and consequently
\begin{multline}
\LL_{\vec{M}}(\scal{\alpha}{\beta})\eps={\LL_{\vec{M}}\alpha}\we*{\beta}+{\alpha}\we*{\LL_{\vec{M}}\beta}-\\-(\vth^{B}\lr\LL_{\vec{M}}\theta^A)\Big((\vth_A\lr\alpha)\we*(\vth_B\lr\beta)+(\vth_B\lr\alpha)\we*(\vth_A\lr\beta)\Big)={\LL_{\vec{M}}\alpha}\we*{\beta}+{\alpha}\we*{\LL_{\vec{M}}\beta}-\\-\LL_{\vec{M}}\theta^A\we\vth^B\lr\Big((\vth_A\lr\alpha)\we*(\vth_B\lr\beta)+(\vth_B\lr\alpha)(\vth_A\lr\beta)\Big).
\label{L-scal}
\end{multline}
The Lie derivative of the volume form $\eps$ can be calculated as follows
\begin{multline*}
\LL_{\vec{M}}\eps=d(\vec{M}\lr\eps)=d\Big(\sum_{a=1}^n(-1)^a\sqrt{(-1)^m\det q_{ij}}M^a dx^1\we\ldots dx^{a-1}\we dx^{a+1}\we\ldots\we dx^n\Big)=\\=
\sum_{a=1}^n(-1)^a\partial_b(\sqrt{(-1)^m\det q_{ij}}M^a)dx^b\we dx^1\we\ldots dx^{a-1}\we dx^{a+1}\we\ldots\we dx^n=\\=\partial_a(\sqrt{(-1)^m\det q_{ij}}M^a)dx^1\we\ldots\we dx^n.
\end{multline*}
Using \eqref{psdq-pt} we calculate further 
\begin{multline*}
\partial_a(\sqrt{(-1)^m\det q_{ij}}M^a)=\frac{\partial \sqrt{(-1)^m\det q_{ij}}}{\partial \theta^A_c}M^a\partial_a\theta^A_c+\sqrt{(-1)^m\det q_{ij}}\partial_aM^a=\\=\sqrt{(-1)^m\det q_{ij}}\,\theta^c_A(M^a\partial_a\theta^A_c+\theta^A_a\partial_cM^a)=\sqrt{(-1)^m\det q_{ij}}\,\vth_A\lr\LL_{\vec{M}}\theta^A, 
\end{multline*}
hence
\[
\LL_{\vec{M}}\eps=\eta_{AB}(\vth^B\lr\LL_{\vec{M}}\theta^A)\eps
\]
and
\[
\scal{\alpha}{\beta}\LL_{\vec{M}}\eps=\LL_{\vec{M}}\theta^A\we\vth^B\lr(\eta_{AB}\alpha\we*\beta).
\]
The above equation and \eqref{L-scal} give the desired identity \eqref{L-*} for $k>0$.

Consider now the case $k=0$. For this value of $k$ the last term disappears from the r.h.s. of \eqref{L-sc-ab}. Consequently, the last term disappears from the r.h.s. of \eqref{L-scal} and thus
\[
\LL_{\vec{M}}(\alpha\we *\beta)={\LL_{\vec{M}}\alpha}\we*{\beta}+{\alpha}\we*{\LL_{\vec{M}}\beta}+\LL_{\vec{M}}\theta^A\we\vth^B\lr(\eta_{AB}\alpha\we*\beta).
\]  
But for zero-forms $\alpha,\beta$ the r.h.s of this formula coincides with the r.h.s. of \eqref{L-*} which means that \eqref{L-*} holds for every $k\in\{0,1,\ldots,n\}$.    
\end{proof}

\end{document}